\documentclass[10pt,journal,compsoc]{IEEEtran}

\usepackage{graphicx} 
\usepackage{booktabs} 
\usepackage{tikz}
\usepackage{tikzscale}
\usepackage{bbding}
\usepackage{amssymb}
\usetikzlibrary{calc}
\usetikzlibrary{decorations.pathmorphing}
\usepackage{forest}
\usepackage{algorithm} 
\usepackage{algorithmicx} 
\usepackage{algpseudocode} 
\usepackage{amsmath} 
\usepackage{graphicx}
\usepackage{subcaption}
\usepackage{cancel}
\usepackage{lipsum}
\usepackage{threeparttable}
\usepackage{color}
\usepackage{multirow} 
\usepackage{graphicx} 
\usepackage{array} 
\usepackage{tabularx}
\usepackage{hyperref}
\usepackage{pgfplots}
\pgfplotsset{compat=1.18}
\ifCLASSOPTIONcompsoc
  \usepackage[nocompress]{cite}
\else
  \usepackage{cite}
\fi
\ifCLASSINFOpdf

\else

\fi

\begin{document}

\title{Nezha: Breaking Multi-Rail Network Barriers for Distributed DNN Training}

\author{Enda Yu,
        Dezun Dong,
        Xiangke Liao
\IEEEcompsocitemizethanks{\IEEEcompsocthanksitem The authors are from the College of Computer Science and Technology, National University of Defense Technology. \protect\\
E-mail: \{yuenda,dong,xkliao\}@nudt.edu.cn
}
\thanks{Dezun Dong is the Corresponding Author.}}

\markboth{arxiv, 2025}%
{Shell \MakeLowercase{\textit{et al.}}: Bare Demo of IEEEtran.cls for Computer Society Journals}

\IEEEtitleabstractindextext{%
\begin{abstract}


In distributed deep learning, the communication overhead of allreduce operations remains a critical bottleneck. While modern hardware offers unprecedented performance, over 60\% of production HPC systems still operate on legacy infrastructure (V100 GPUs, multi-plane Ethernet/InfiniBand networks) that requires communication optimization without hardware upgrades. Existing approaches suffer from three limitations: (1) static single-rail binding underutilizes multi-rail bandwidth, (2) protocol heterogeneity (e.g., TCP-RDMA coexistence) causes synchronization delays, and (3) mainstream communication libraries (NCCL/MPI) lack cross-protocol coordination. We present Nezha, the first protocol-agnostic allreduce system for multi-rail networks. Our key innovations include: \textbf{Hardware-agnostic cross-protocol coordination}: A unified abstraction layer enabling seamless collaboration between in-network computing (SHARP), adaptive RDMA (GLEX), and TCP, achieving 1.7–4.3× lower latency than Gloo.
 \textbf{Protocol-aware dynamic load balancing}: A hybrid scheduling strategy incorporating a cold/hot start state machine for heterogeneous protocols (TCP, SHARP, GLEX), which reduces startup latency for small payloads and enhances throughput for large-scale transfers.
 \textbf{Fault-tolerant multi-rail collaboration}: A self-recovery mechanism rerouting data flows within 200 ms upon single-rail failures, ensuring uninterrupted training. 
Experiments on 8-node clusters show Nezha achieves 74\% and 80\% higher throughput than MPTCP in homogeneous (TCP-TCP) and heterogeneous (TCP-SHARP) networks, respectively. Furthermore, Nezha delivers 2.36x higher training efficiency than Gloo on 128-node supercomputer systems. Nezha bridges the gap between modern DNN communication demands and legacy infrastructure, demonstrating that systematic multi-rail optimization can unlock the potential of aging clusters.

\end{abstract}

}

\maketitle
\IEEEdisplaynontitleabstractindextext
\IEEEpeerreviewmaketitle

\section{Introduction}

\begin{figure}[!t]
  \centering
  \resizebox{\linewidth}{!}{\includegraphics{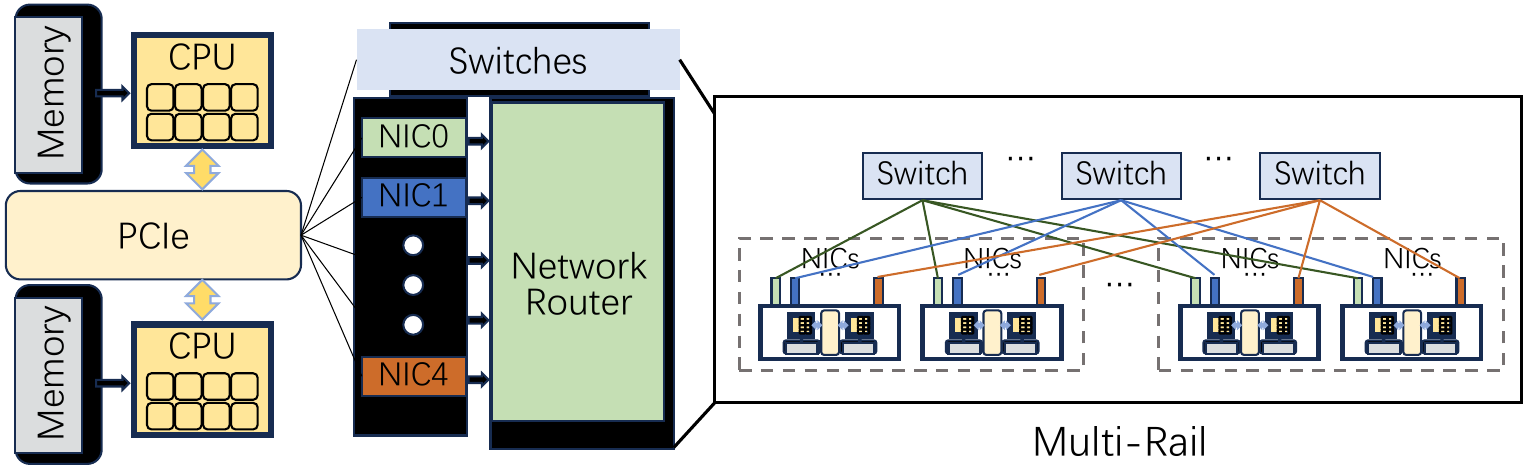}}
  \caption{Multi-rail networks architectures.}
  \label{duogui}
\end{figure}

Distributed deep learning has become the cornerstone of modern AI breakthroughs, with data-parallel training paradigms \cite{yu2023communication,ddp,fsdp} relying on efficient allreduce operations to synchronize gradients across thousands of parameters \cite{gibiansky2017bringing,de2024swing}. While cutting-edge hardware (e.g., NVLink \cite{nvlink}, B100 GPUs \cite{bgpu}) has significantly accelerated distributed deep learning, over 60\% of Chinese production HPC systems \cite{top500} still rely on legacy infrastructures due to lengthy upgrade cycles and the semiconductor export administration regulations (EAR) \cite{ear}. Although newer high-performance network devices are gradually being deployed in these environments, legacy equipment is often retained in service owing to budget constraints, resulting in heterogeneous multi-plane network topologies (e.g., dual-rail InfiniBand (IB), hybrid Ethernet/IB configurations) rather than homogeneous GPU-centric interconnects. This coexistence of old and new hardware creates physically isolated network paths for data transmission, forming multi-rail network architectures as depicted in Fig.\ref{duogui}. Specifically, servers frequently integrate multiple generations of NICs (e.g., 100Gbps IB HCA and 40Gbps Ethernet adapters) to maximize resource utilization, further amplifying protocol heterogeneity across network planes. Such configurations are prevalent in China's state-sponsored supercomputing centers and data centers, where geopolitical constraints and cost-effectiveness jointly drive the adoption of incremental hardware upgrades. However, our experimental analysis on an 8-node production cluster reveals that existing solutions \cite{ecf,nvidia2017nccl} achieve only 41.2\%±9.7\% of theoretical multi-rail bandwidth during allreduce operations. This indicates a systemic underutilization crisis given the exponential growth of DNN model sizes.


This inefficiency stems from three fundamental limitations in current approaches: 1. \textbf{Static single-rail binding}: Training frameworks \cite{paszke2019pytorch,abadi2016tensorflow} default to the lowest-latency single link, leaving 30-50\% of the available bandwidth in multi-rail networks idle during peak computation phases \cite{jiang2024megascale}. This coarse-grained network selection leads to systemic hardware underutilization. 2. \textbf{Protocol heterogeneity blindness}: Existing multi-path solutions such as MPTCP \cite{ecf} (designed for general-purpose data transfer) and MRIB \cite{2004Building} (originally targeting traffic engineering in wide-area networks) are limited to homogeneous link assumptions, neglecting the unique protocol heterogeneity in distributed DNN training. For instance, their static path scheduling fails to address the delay imbalance caused by in-network computing protocols (e.g., SHARP’s switch-level aggregation \cite{graham2016scalable}) and RDMA’s out-of-order execution. Consequently, synchronization delays from protocol mismatch lead to 15-40\% throughput degradation in hybrid networks. 3. \textbf{Hardware-coupling pitfalls}: Vendor-specific libraries like NCCL \cite{nvidia2017nccl} and RCCL \cite{rccl}) achieve peak performance on target architectures \cite{nvidia,amd} but suffer catastrophic failures (up to 10× slowdown) on mixed-vendor clusters, while providing no cross-protocol coordination mechanisms.

These limitations reveal a critical conflict: the growing communication efficiency demands of modern models versus the inefficient utilization of basic network infrastructure in HPC facilities. Traditional approaches treat multi-rail networks as homogeneous bandwidth pools, ignoring three unique challenges exposed in our study: (a) non-linear latency composition when merging in-network computing paths with traditional transports, (b) dynamic load imbalance from protocol-specific communication patterns, and (c) fault propagation risks in cross-protocol synchronization.

To address these challenges, we present Nezha\footnote{\label{footnote:Nezha}Available at: \url{https://github.com/nezha-Anonymous/nezha}}, the first protocol-agnostic allreduce system that dynamically coordinates heterogeneous network planes through adaptive tensor partitioning and hardware-oblivious fault tolerance. Nezha introduces the following key innovations:




\noindent$\bullet$ \textbf{Hardware-agnostic cross-protocol coordination:}
A unified abstraction layer enabling simultaneous utilization of in-network computing SHARP \cite{graham2016scalable}, adaptive RDMA GLEX \cite{15-2-5-4603}, and conventional TCP through dynamic plane orchestration, achieving 1.7-4.3× lower allreduce latency than Gloo in multi-rail networks.

\noindent$\bullet$ \textbf{Protocol-aware dynamic load balancing:}
A hybrid scheduler dynamically partitions data across heterogeneous links. For small payloads, Nezha operates in a \textbf{cold start} state, routing all data through the lowest-latency protocol. For large transfers, it transitions to a \textbf{hot start} state, balancing load based on real-time latency feedback.

\noindent$\bullet$ \textbf{Fault-tolerant multi-rail collaboration:}
A self-recovery mechanism that automatically detects network failures and switches to backup channels, ensuring continuous operation through seamless protocol transition.

Our evaluation on 8-node clusters demonstrates Nezha's effectiveness: it achieves 58-87\% throughput gains in homogeneous dual-rail TCP networks and 41-63\% improvements in heterogeneous TCP-SHARP/GLEX configurations, while reducing communication startup overhead by 15-23\% compared to MPTCP and MRIB. When used as a communication backend for GPT-3 \cite{gpt} training on 128 nodes, Nezha achieves 2.36x higher training efficiency than Gloo in homogeneous multi-rail networks.

\section{Background}

This section analyzes communication bottlenecks in distributed DNN training and explores the limitations of multi-rail network utilization. We first use theoretical modeling and empirical measurements to identify the root causes of these bottlenecks. Then, we evaluate existing network enhancement approaches from two aspects: multi-rail bandwidth aggregation and hardware acceleration. Finally, through empirical analysis of production clusters, we identify three key challenges in deploying distributed communication operations within multi-rail network architectures.

\subsection{Communication Bottlenecks in Distributed Systems}
\label{sec:comm_bottlenecks}

Modern distributed systems face fundamental communication constraints that hinder training efficiency. As model sizes grow exponentially, three critical bottlenecks emerge in parallel training and inference tasks.

\subsubsection{Cross-paradigm communication overheads}
The allreduce operation exhibits distinct communication characteristics across parallelization strategies \cite{shoeybi2019megatron}. In data parallelism implementations such as PyTorch DDP \cite{ddp}, allreduce synchronizes gradients through ring topology with communication complexity $C$ determined by:

\begin{equation}
C = 2(N-1)\frac{M}{N}
\end{equation}

where $M$ denotes gradient volume and $N$ denotes the number of workers. Practical deployments reveal prohibitive memory demands: training GPT-2 (1.5B parameters) \cite{gpt} on 16 V100 GPUs requires synchronizing 11.25 GB gradients per iteration, consuming 35\% of available GPU memory. While techniques like DeepSpeed-ZeRO \cite{deepspeed} mitigate memory pressure through phase decomposition, they increase total communication volume by 1.5$\times$. Model parallelism introduces complexity through allreduce operations across layers. For example, GPT-3 executes 12 allreduce operations per attention layer, each exchanging 603 MB of data. Tensor parallelism further amplifies demands: deploying LLaMA-3 (70B parameters) \cite{llama} on 4 L40 GPUs requires 160 allreduce operations in each inference iteration, contributing over 50\% of the prefill cost.

\subsubsection{Single-rail bandwidth wall}
\label{subsec:scalability}

Distributed systems face significant scalability limitations due to the inefficiencies of single-rail allreduce operations. For example, when training VGG-16 \cite{VGG} on 16 V100 GPUs, communication overhead accounts for 72\% of the total iteration time (957 ms out of 1329 ms), primarily due to insufficient inter-node bandwidth. This challenge can be formalized using the network efficiency model:

\begin{equation}
\label{eq2}
\delta_{\text{net}}= \frac{\hat{T}_{\text{theoretical}}}{T_{\text{actual}}}  =\frac{1}{1 +  \frac{T_{\text{setup}}}{S/B}}
\end{equation}

Here, $S$ denotes the message size, $B$ represents the effective link bandwidth, $T_{\text{setup}}$ encapsulates fixed per-message latency components (including protocol processing and queuing delays). \(\hat{T}_{\text{theoretical}}\) represents the ideal transmission time without any overheads, while \(T_{\text{actual}}\) includes all real-world latency components. This formulation reveals a critical dependence on message granularity: $S/B$ captures the theoretical transmission time, while $T_{\text{setup}}$ incorporates the fixed protocol overhead that becomes dominant for small messages. This explains the low network efficiency observed in small-scale data allreduce operations, where small payloads spend a disproportionate amount of time on protocol overhead rather than productive data transfer. Consequently, the network optimization design of distributed systems must carefully consider communication granularity to reduce transmission time and mitigate the impact of fixed latency.

\subsubsection{Hardware disparity and fundamental limits}
\label{subsec:hardware}

The growing disparity between computational throughput and network bandwidth imposes fundamental architectural constraints. From V100 to H100, NVIDIA's FP16 arithmetic throughput $\mathcal{F}_{\text{GPU}}$ increased 7.9×, while network interface bandwidth $B_{\text{phy}}$ improved only 4×.  Additionally, the bottleneck coefficient \(\beta = \mathcal{F}_{\text{GPU}} / \mathcal{B_{\text{phy}}}\) shows that modern GPUs outpace network capabilities by over an order of magnitude. These results demonstrate that single-link architectures cannot maintain equilibrium, necessitating innovations in multi-rail bandwidth aggregation and hardware-assisted acceleration.

\subsection{Communication Optimization in Distributed Systems}

Modern distributed systems employ two dominant approaches to enhance network performance: multi-rail bandwidth aggregation and hardware-assisted acceleration. Our comparative analysis reveals their complementary strengths and inherent limitations.

\subsubsection{Multi-rail network technologies}

Current multi-rail network technologies are primarily designed for general data transmission, multimedia transmission, and web-based applications, rather than specifically for distributed deep learning. Consequently, when integrated with allreduce operations, several limitations arise. Standardized protocols like MPTCP \cite{he2023survey,barre2011multipath,ecf} aggregate bandwidth through multiple TCP subflows but exhibit inefficiencies in heterogeneous settings due to their reliance on round-trip time (RTT) based scheduling. While enhancements such as the ECF \cite{ecf} scheduler improve bandwidth utilization by dynamically selecting paths based on bandwidth estimation and completion time prediction, they cannot understand the completion time differences between heterogeneous protocols, which can cause significant synchronization delays in distributed deep learning.

Similarly, MRIB \cite{2004Building} achieves hardware-aware bandwidth aggregation using multiple IB HCAs, ports, or virtual paths controlled by LID masks. Its virtual subchannel abstraction maps physical links to logical communication channels, and sets data allocation weights based on the bandwidth of each communication channel. In response to network fluctuations or sudden congestion, it dynamically adjusts weights according to the current transmission delay differences across communication channels to enhance bandwidth utilization. However, MRIB's strategy of seeking a static optimal weight allocation to maximize bandwidth efficiency has limitations in distributed training. This is because the gradient data packets generated by nodes vary in size, leading to dynamic changes in network efficiency of heterogeneous protocols (Eq. \ref{eq2}), which contradicts the assumption of static weight optimization.

\subsubsection{RDMA-based network protocols}

Modern distributed systems leverage RDMA-based protocols to mitigate communication bottleneck. These protocols primarily consist of two paradigms: in-network computing and customized interconnects in supercomputing environments. SHARP, an in-network computing protocol, reduces end-host processing by performing switch-level aggregation, achieving 1.8-3.2× throughput improvements for medium-sized tensors. High-performance computing systems use customized interconnect protocols like Fugaku's TofuD \cite{tofu}, Sunway's SWverbs \cite{gao2021interconnection}, and TH Express-2's GLEX \cite{15-2-5-4603} to optimize collective operations. These protocols demonstrate that hardware-assisted optimizations can significantly enhance communication efficiency, yet their effectiveness varies across different data sizes and workloads.


\begin{figure}[!t]
  \centering
  \resizebox{\linewidth}{!}{\includegraphics{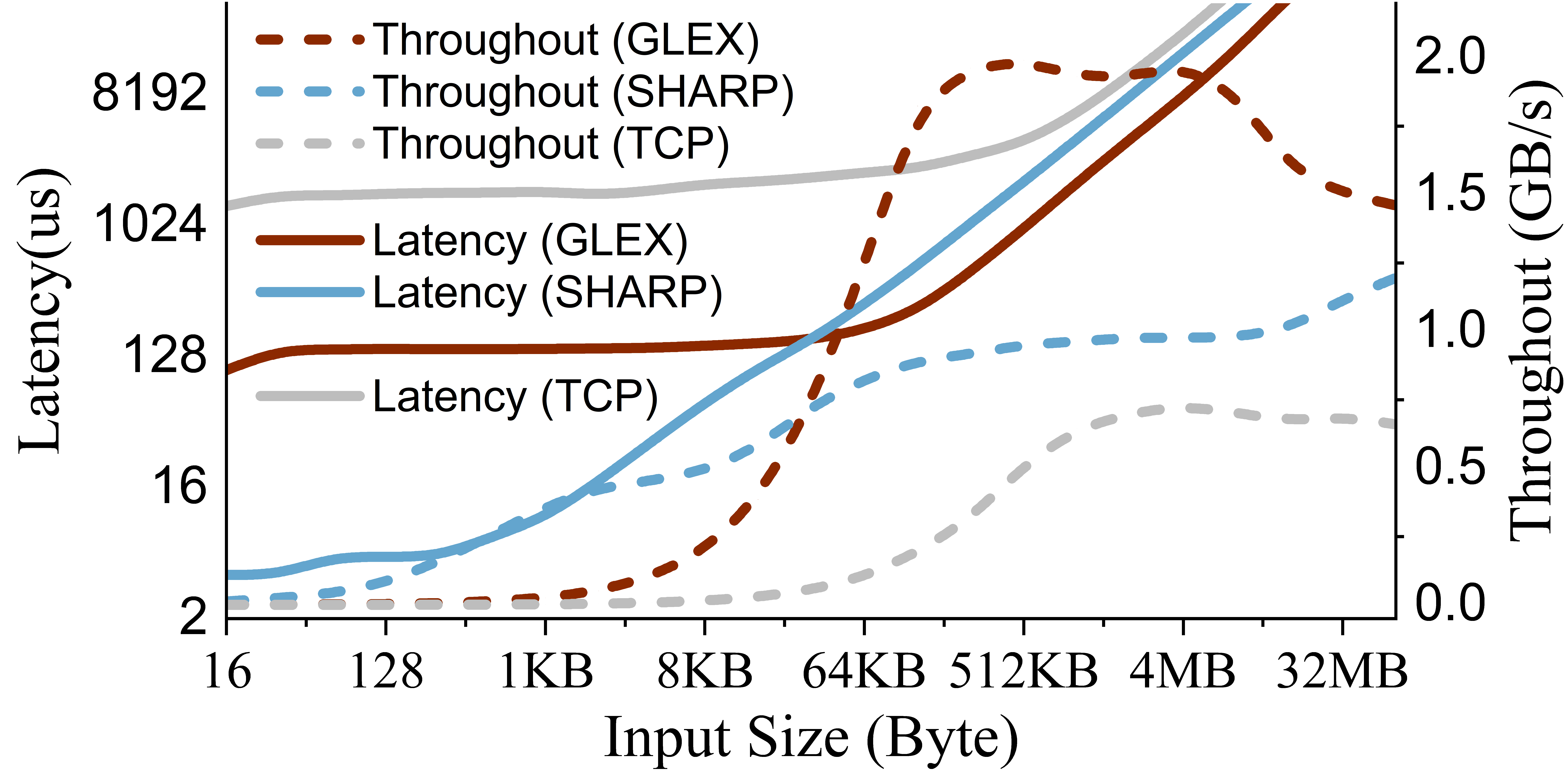}}
  \caption{Latency and throughput characteristics of GLEX, TCP, and SHARP protocols in allreduce operations across varying data sizes.}
  \label{tsg}
\end{figure}

\subsection{Key Challenges in Multi-Rail Networks}
While subsection \ref{sec:comm_bottlenecks} identifies the fundamental communication bottlenecks in distributed systems, multi-rail networks introduce three unique challenges that further exacerbate these limitations in production environments. Our analysis of 8-node production clusters reveals three critical gaps between theoretical network capabilities and practical performance.

\subsubsection{Protocol Heterogeneity Blindness}

Current multi-path protocols assume homogeneous environments, conflicting with the heterogeneous protocols in distributed systems. However, in China, funding constraints and slow equipment renewal lead over 60\% of data centers and HPC systems to retain older network equipment alongside new devices. Examples like Paratera Cloud Cluster and the Tianjin Data Center, which support both Ethernet and IB networks, and TH Express-2, which uses Ethernet and TH networks, highlight the challenges of utilizing heterogeneous multi-rail networks to support distributed systems. As shown in Fig. \ref{tsg}, the protocol throughput varies significantly with data size: SHARP achieves ultra-low latency for small messages (under 256KB), while GLEX maintains higher throughput for payloads of 64KB-64MB. Customized optimization causes non-linear changes in network throughput, rendering MRIB's static bandwidth-based weight allocation. Additionally, cross-protocol asymmetry in bandwidth-delay product exacerbates inefficiencies. For example, SHARP achieves 0.73 GB/s effective bandwidth at 32KB, whereas TCP delivers only 0.06 GB/s. Under such conditions, MPTCP's RTT-driven data slicing strategy causes TCP links to become systemic bottlenecks.
To address these challenges, we propose \textit{the first design proposition}: data partitioning across rails is only activated when the real-time efficiency ratio $\rho(S)$ between networks does not exceed the protocol divergence tolerance threshold $\tau$, formulated as:

\begin{equation}  
\label{abs}
\rho(S) = \frac{S^i \cdot \delta_{\mathrm{i}}/\hat{T}_i}{S^j \cdot \delta_{\mathrm{j}}/\hat{T}_j} \quad\quad S  = S^i + S^j
\end{equation}
The numerator and denominator represent the real-time throughput of network i and network j when processing the data volume of $S^i$ and $S^j$, respectively. Fig. \ref{rho} shows the throughput improvement ratio that the optimal network can achieve in the ideal state among multi-rail network combinations with different real-time efficiency ratios. In the case of $\rho(S)=1 $, the optimal improvement can be obtained, which means that the multi-rail network is 
homogeneous. When $\rho(S)$ is greater than 5, the throughput revenue tends to slow down. After our testing, setting the tolerance threshold $\tau$ to 5 is the most reasonable. At this point, the theoretical throughput revenue will be offset by the negative effects of synchronization overhead and system resource contention.

\begin{figure}[!t]
  \centering
  \resizebox{\linewidth}{!}{\includegraphics{quan.tikz}}
  \caption{The impact of real-time efficiency ratio on the throughput improvement ratio of the optimal network.}
  \label{rho}
\end{figure}

\subsubsection{System Resource Contention}
\label{cpure}
\begin{figure}[!t]
  \centering
  \resizebox{\linewidth}{!}{\includegraphics{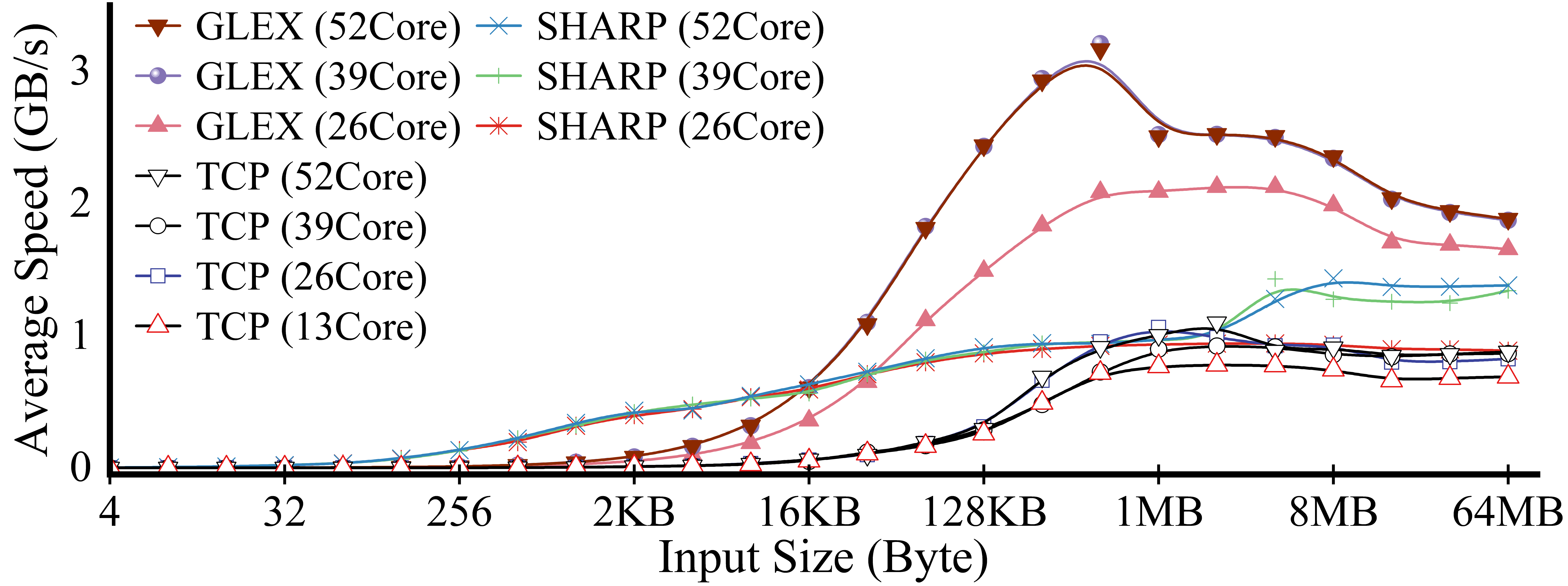}}
  \caption{Throughput of allreduce on various single-rail networks bound with different CPU cores.}
  \label{cpu}
\end{figure}

The interplay between protocol-specific resource demands and heterogeneous network architectures generates critical contention points in multi-rail systems, as demonstrated by the throughput-core relationships in Fig.\ref{cpu}. GLEX and SHARP exhibit strong scalability with increased CPU allocation, reflecting their reliance on CPU resources for protocol control mechanisms. Conversely, TCP allreduce operations show insensitivity to CPU core count scaling (reaching peak throughput with only 26 cores). This divergence stems from fundamental architectural differences: GLEX offloads data movement to RDMA while retaining CPU-intensive control-plane operations such as queue management. In contrast, SHARP reduces end-host computation through in-network aggregation but incurs metadata synchronization overhead. When co-deployed, these protocols compete for shared resources—as exemplified by dual-rail GLEX+TCP configurations with equal core allocation (26 cores per protocol), achieving only 68\% of their combined peak throughput.

This contention invalidates traditional equal-partitioning strategies. Allocating equivalent resources across heterogeneous rails, such as distributing 26 cores equally among GLEX, SHARP, and TCP, causes severe performance degradation. Specifically, the throughput of SHARP and GLEX decreases by 42\% and 35\%, respectively, compared to their peak throughput. Such penalties arise because static partitioning cannot reconcile protocol-specific resource profiles under dynamic allreduce workloads. To address this, we propose \textit{the second design proposition}: adaptive dynamic resource partitioning that allocates compute resources proportionally to runtime protocol requirements during allreduce operations.

\begin{figure*}[!t]
  \centering
  \resizebox{\linewidth}{!}{\includegraphics{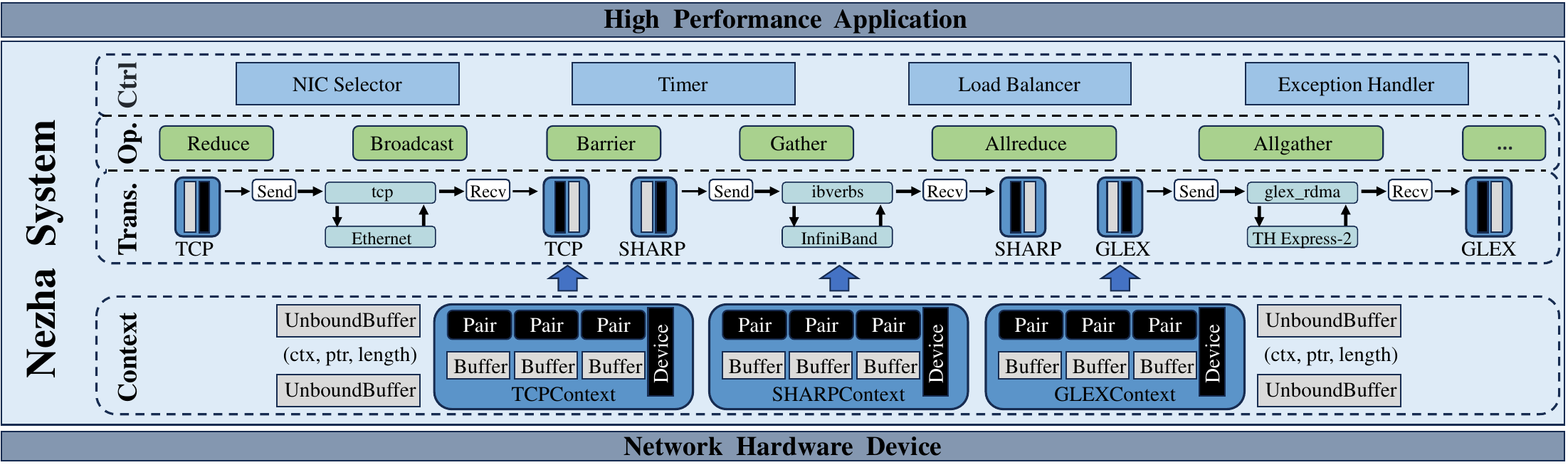}}
  \caption{The system architecture of Nezha. Context, Trans., Op., and Ctrl represent the Context, Transport, Collective Operations, and Control modules respectively.}
  \label{jiagou}
\end{figure*}
\subsubsection{Emergency Fault Handling}

The operational resilience of heterogeneous multi-rail networks is significantly compromised due to inadequate fault tolerance capabilities, which disrupt task continuity in production environments. Empirical studies have identified three primary failure modes: First, hardware vulnerabilities under thermal stress conditions pose a critical challenge. For example, 85\% of GPU temperature elevation incidents trigger power-off protection in the IB network interface when cluster temperature control systems fail to perform adequately. This results in an average 23-minute training suspension in an 8-node cluster during re-discovery and connection recovery. Second, protocol-induced fault propagation arises from the complexity of network transport protocols. This complicates configuration management and can lead to performance degradation or connection failures, further exacerbated by compatibility issues between multi-vendor network equipment. Third, operational blind spots in fault management, such as the lack of visualization tools and centralized platforms, delay root cause identification and increase recovery time. For instance, MPTCP-induced uncoordinated failover has been shown to cause 21.3\% gradient divergence and 15-minute training interruptions in experiments. To address these challenges, we propose \textit{the second design proposition}: heterogeneous multi-rail networks should establish fault isolation boundaries between each other and establish a task migration mechanism to surviving networks during sudden network failures.


\section{Nezha System}

This section introduces the system design of Nezha, with a focus on its key design objectives and architecture.

\subsection{Overview}\label{AA}

Nezha uses a modular architecture to support different network protocols. In addition to the TCP protocol, Nezha integrates two RDMA-based protocols: SHARP, a popular in-network computing protocol, and GLEX, chosen from custom networks. The system architecture, depicted in Fig.\ref{jiagou}, has four key modules. The Context Module manages global communication context. The Transport Module handles node-to-node communication. The Collective Module implements collective communication operations. The Control Module coordinates multi-rail network collaboration and ensures reliable data transmission.

\subsection{Context Module}
The Context Module provides unified interfaces (TCPContext, SHARPContext, GLEXContext) for different protocols. Each context object uses \textup{Pair} objects for point-to-point communication and manages private resources such as NIC device binding and buffer allocation. To achieve efficient data processing, a cross-protocol shared buffer mechanism is designed. Data is initially placed in \textup{UnboundBuffer} before allocation. Each \textup{Pair} uses \textit{ptr} and \textit{data\_length} to indicate the reading position. Pairs can asynchronously transfer data from \textup{UnboundBuffer} to \textup{Buffer} for processing, and then return the results to \textup{UnboundBuffer}. Once all results are collected, \textup{UnboundBuffer} makes the data available to the data requester and is subsequently destroyed. The Context Module also integrates the necessary functional modules of customized networks, such as SHARP's aggregation tree and GLEX's memory registration module.

\subsection{Transport Module}
The Transport Module employs the rendezvous mechanism to establish global communication connections. In addition to \texttt{tcp}, it also supports \texttt{ibverbs} and \texttt{glex\_rdma} for implementing point-to-point communication between \textup{Pairs}. The \texttt{ibverbs} segment is tailored for SHARP, verifying the creation of the collective communication domain and SHARP tree to enable in-network computing capabilities. For \texttt{glex\_rdma} transmission, additional data structures such as \texttt{send\_req} and request queues \texttt{send\_reqs} are required. In cases where operations within \textup{Buffer} fail to complete in a timely manner, the initiating memory address, communication sequence number, and an uncompleted flag are stored in \texttt{send\_req}, which is then placed into request queues \texttt{send\_reqs}. Both sending and receiving entities monitor \texttt{send\_reqs} for pending tasks, ensuring non-blocking operations between \textup{Pairs} and enhancing communication efficiency.

\subsection{Collective Operations Module}

The Collective Operations Module provides implementation for a variety of collective operations, to maintain clarity and relevance to our central thesis, this paper focuses exclusively on the allreduce operation. Each type of collective communication operation corresponds to a derived class. Each class possesses an operational handle, Opts, utilized to identify the received context object, thereby determining the transmission mode for each operation. Moreover, Opts provides an interface (\textit{ptr},\textit{data\_length}), where \textit{ptr} is a pointer to the starting address of the data to be processed, and \textit{data\_length} is the length of the data. Each network retrieves the data based on the received (\textit{ptr},\textit{data\_length}) from Opts, ensuring collaborative data read and write operations among networks.


\begin{figure}[t!]
  \centering
  \resizebox{0.8\linewidth}{!}{\includegraphics{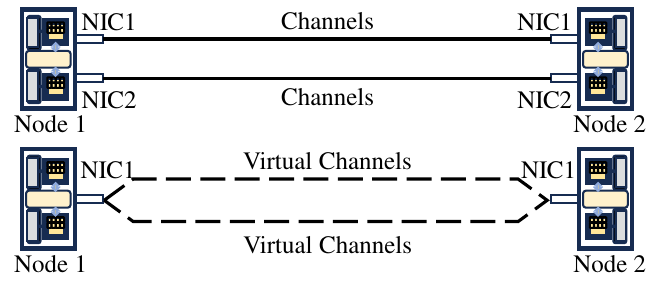}}
  \caption{Channel abstraction of Nezha.}
  \label{yingjian2}
\end{figure}
\subsection{Control Module}
The Control Module comprises several critical components, including NIC Selector, Timer, Load Balancer, and Exception Handler. The NIC Selector responds to each constituent network participating in multi-rail networks, selects appropriate devices for each network, and creates corresponding context objects. The Timer monitors the latency of each network at the onset of collective communication and transmits the collected data to the Load Balancer. The Load Balancer designs the data allocation scheme based on the data from the Timer, adjusts the volume of data received by each network adapter, and calculates the pointer addresses \textit{ptr} used by each network thread for data processing based on the data volume (section \ref{dataallsc}). Meanwhile, the Exception Handler oversees network status and data synchronization assurances, and enhances the network's fault tolerance (section \ref{exce}).

\section{Multi-Rail Networks on Nezha}

This section introduces how Nezha supports allreduce on multi-rail networks, including channel abstraction, the working process of each module, cross-network data allocation schemes, and fault-tolerant design.


\subsection{Channel Abstraction of Nezha}
\label{channel}

Nezha ensures efficient data forwarding in multi-rail networks by providing dedicated support for each member network, preventing cross-network interference. As shown in Fig.\ref{yingjian2}, in clusters with multiple NICs per node, Nezha assigns separate channels on each physical link, each exclusively used by a member network thread. Moreover, Nezha can serve nodes without multiple NICs by creating multiple virtual channels on the same physical link when permitted by the protocol. These virtual channels are then used by member networks to perform allreduce operations, forming what we call a virtual multi-rail network. This approach allows Nezha to efficiently utilize hardware resources and enhance network performance and resource usage.


\begin{figure}[!t]
  \centering
  \resizebox{\linewidth}{!}{\includegraphics{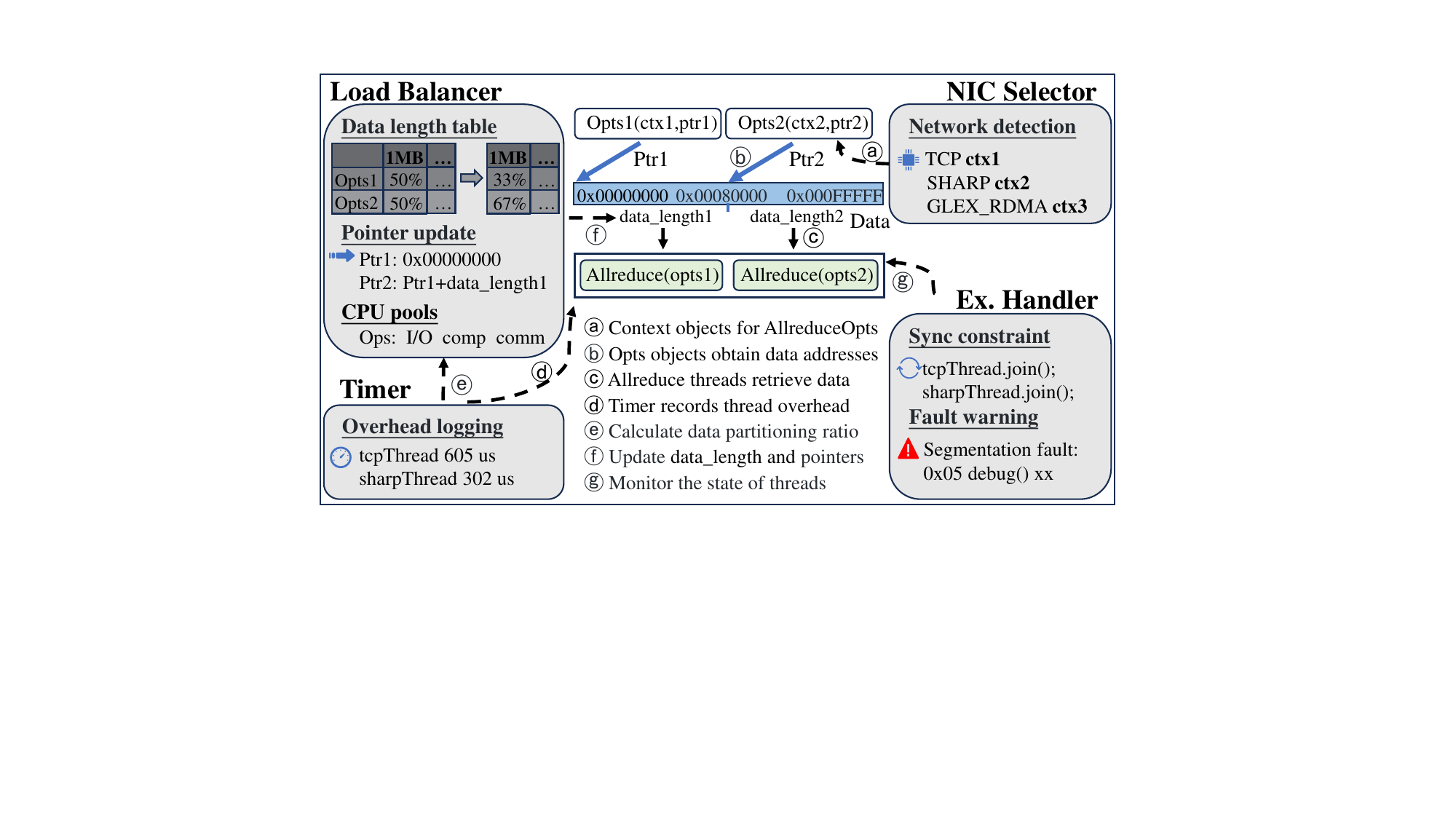}}
  \caption{The process of allreduce on multi-rail networks in Nezha system.}
  \label{fenge}
\end{figure}

\subsection{Allreduce on Multi-Rail Networks}

The process of allreduce in the Nezha system is depicted in Fig.\ref{fenge}. During the initialization phase, the NIC Selector creates context objects for networks participating in allreduce. Subsequently, each network's Opts retrieve the memory addresses of the data using pointers. The allreduce threads then access the Opts, read the data they are responsible for, and initiate data transmission. As communication commences, the Timer records the cost of each operation based on the thread ID. To mitigate decision errors stemming from fluctuating costs, the Timer transmits the average cost of every 100 allreduce operations with the same data size to the Load Balancer. Guided by cost optimization, the Load Balancer adjusts data volume and pointers for each network. It maintains a data length table reflecting each member network’s responsibility ratio for different data sizes. The Load Balancer allocates system resources based on CPU pools, dividing the allreduce task into three phases: data loading (I/O), cross-node transfer (communication), and data aggregation (computation). Only the computation phase requires sufficient CPU cores, with most cores released in other phases. It first determines the maximum CPU cores available to the participating multi-rail network members, then dynamically allocates and recovers cores during allreduce based on thread phases. The Exception Handler receives system failure notifications. In case of network failure, it records the faulty network object and triggers the exception handling scheme.

\subsection{Data Allocation Scheme}
\label{dataallsc}

\begin{table}[t!]
\caption{Average latency $T$ of allreduce on 4 nodes (us).}
\begin{center}
\resizebox{0.9\linewidth}{!}{%
\begin{tabularx}{\linewidth}{ c|c c c c c c  }
{Data $S$}&{SHARP} &{TCP} &{T/S$^{1/1}$} &{T/S$^{99/1}$} &{T/S$^{1/99}$}& T/S$^{slic}$ \\
\cline{1-7} 
{1KB}&9 &982 &987 &984 &991&1002 \\
\cline{1-7} 
{8MB}&22140 &37137 &21265 &37141 &23911&31013 \\
\cline{1-7} 
{64MB}&181484 &316323 &178373 &314913 &188137&257135 \\
\cline{1-7}
\multicolumn{7}{X}{T/S$^{x/y}$ denotes x\% data to TCP and y\% to SHARP. T/S$^{slic}$ adopts MPTCP's data slicing strategy.}
\end{tabularx}
\label{tab1}
}
\end{center}
\end{table}

Prior to formulating our data allocation scheme, we conducted systematic evaluations on TCP-SHARP multi-rail networks, comparing the performance of two mainstream approaches: MRIB's fixed-ratio allocation and MPTCP's packet slicing strategy. Through 10,000 allreduce operations per data scale on a 4-node cluster, we obtained average latency metrics as shown in Table \ref{tab1}. Experimental analysis reveals three key findings: First, the real-time efficiency ratio $\rho(S)$ (Eq. \ref{abs}) between heterogeneous networks demonstrates significant data-scale sensitivity. When $\rho(S)$ exceeds the protocol divergence tolerance threshold $\tau=5$, no allocation strategy can provide performance benefits for multi-rail networks, necessitating dynamic decision control through a critical threshold $S_{\text{threshold}}$. Second, protocol synchronization penalties and slice metadata overhead make frequent cross-network partitioning strategies suboptimal in heterogeneous environments, with MPTCP slicing introducing 18-27\% additional latency. Third, resource contention and synchronization overhead under load imbalance cause substantial performance degradation, where static allocation schemes exhibit 9\% latency increase compared to dynamic balancing.

Based on these findings, we propose a dual-state transition latency minimization scheme. For small-scale data ($S \leq S_{\text{threshold}}$), the system enters a \textbf{cold start} state, where communication latency follows:
\begin{equation}
T_{\text{cold}} = \min_{i} \left( T_{\text{setup}}^{i} + \frac{S}{B_{i}} \right)
\end{equation}
Here, $T_{\text{setup}}^{i}$ denotes the startup latency of network $i$, and $B_{i}$ represents effective bandwidth. In this state, full data routing to the optimal single network minimizes startup overhead. For large-scale data ($S > S_{\text{threshold}}$), the system switches to a \textbf{hot-start} state, where latency adheres to:
\begin{equation}
T_{\text{hot}} = \max_{i} \left( T_{\text{setup}}^{i} + \frac{\alpha_S^{i} S}{B_{i}} \right)
\end{equation}
with $\alpha_{S}^{(i)}$ indicating the data allocation proportion for network $i$ (satisfying $\sum_i \alpha_{S}^{(i)}=1$). The state transition threshold $S_{\text{threshold}}$ is determined through Eq. \ref{theq}, ensuring latency equivalence at critical points between cold and hot states.
\begin{equation}
\min_{i} \left( T_{\text{setup}}^{i} + \frac{S_{\text{threshold}}}{B_{i}} \right) = \max_{i} \left( T_{\text{setup}}^{i} + \frac{\alpha_S^{i} S_{\text{threshold}}}{B_{i}} \right)
\label{theq}
\end{equation}

Coefficient optimization in the hot-start state employs gradient descent (Eq. \ref{sgd}), iteratively updating $\alpha_{S}^{(i)}$ until data length table convergence.
\begin{equation}
\alpha_S^{i,k+1} = \alpha_S^{i,k} - \eta \frac{\partial T_{\text{hot}}}{\partial \alpha_S^{i,k}} 
\label{sgd}
\end{equation}
To accelerate convergence, the Load Balancer generates initial coefficients $\alpha_{S}^{i,0}$ through Eq. \ref{s0} based on initial uniform allocation where $T$ represents global latency summation, $T_i$ denotes the latency of network $i$ and $N$ is the node count.
\begin{equation}
    \alpha_S^{i,0}=\frac{T-T_i}{T*(N-1)}
    \label{s0}
\end{equation}
Experimental results demonstrate that leveraging temporal stability characteristics of communication patterns in distributed deep learning enables Nezha to complete threshold search and coefficient convergence within the first 100 iterations, achieving sustained minimal latency assurance.



\begin{figure}[!t]
  \centering
  \resizebox{\linewidth}{!}{\includegraphics{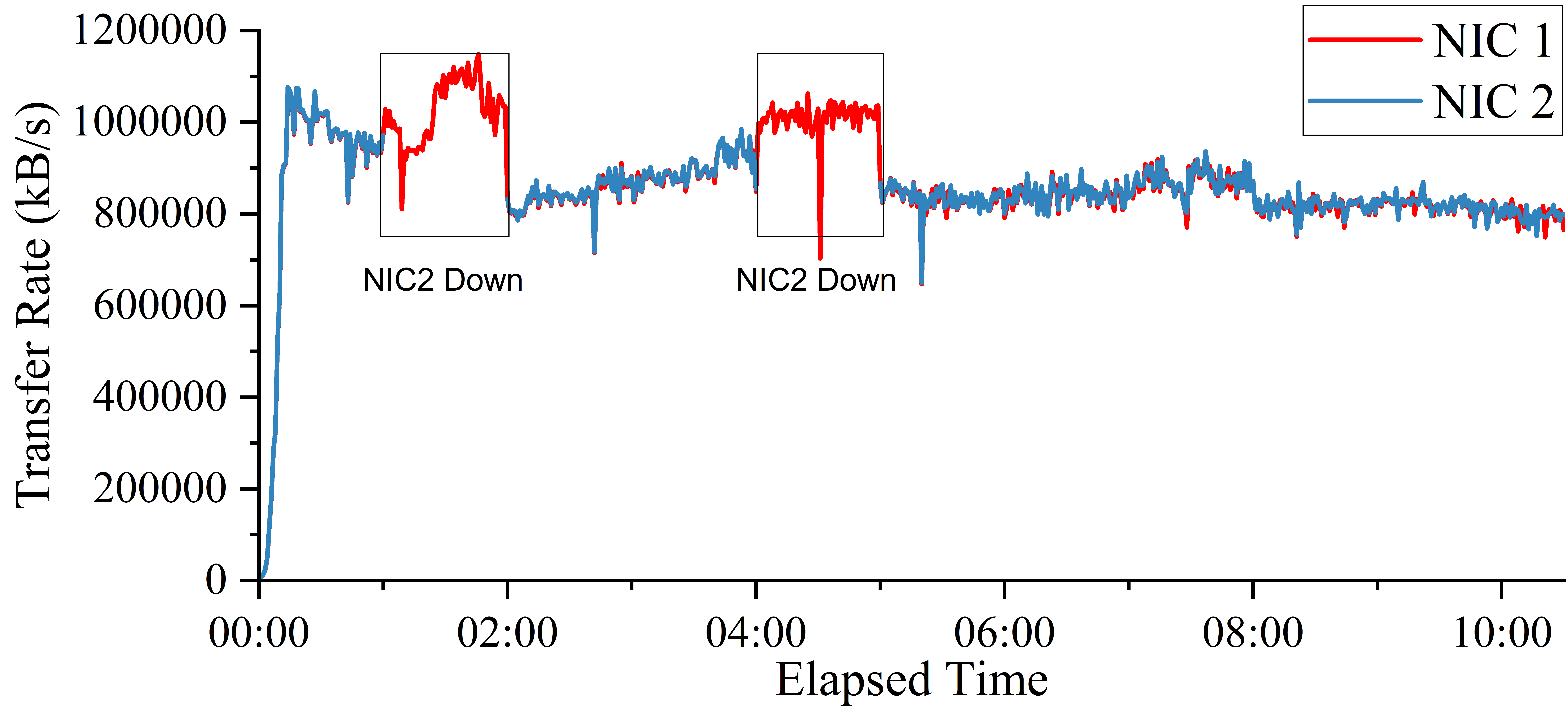}}
  \caption{Average transfer rate of NICs when performing allreduce on dual-rail TCP networks.}
  \label{tran}
\end{figure}

\subsection{Fault Tolerant Network Design}
\label{exce}

 Nezha provides a fault-tolerant network design for multi-rail networks, allowing other healthy member networks to take over tasks from a failed network and continue the training tasks. The workflow of the fault handling scheme is as follows: upon detecting an exception signal from a member network, the Exception Handler sends the fault network's data processing pointer address \textit{ptr} and data length \textit{data\_length} to the unaffected optimal member network and deregisters the fault network's operation handle. The Exception Handler determines the optimal member network based on the data length \textit{data\_length} allocated by the Load Balancer to each network, with the network handling more data typically being more performant. The optimal member network receiving the signal, after completing its current task, passes (\textit{ptr}, \textit{data\_length}) to the operation handle to initiate the allreduce operation.


\begin{figure*}[t!]
 \centering
  \begin{subfigure}[h]{0.48\textwidth}
    \includegraphics[width=\textwidth]{figure/aalatency.png}
    \label{aalatency}
  \end{subfigure}  
  \begin{subfigure}[h]{0.48\textwidth}
    \includegraphics[width=\textwidth]{figure/aaspeed.png}
    \label{aaspeed}
  \end{subfigure}  
  \caption{Latency and throughput of performing allreduce on homogeneous dual-rail TCP network.}
  \label{hetnet}
\end{figure*}

We use the performance analysis tool SAR \cite{sar_manual} to log the transfer rates of the NICs in dual-TCP multi-rail networks during allreduce. The results are shown in Fig. \ref{tran}. In homogeneous TCP-TCP networks, allreduce data is equally divided between the two NICs. This results in their respective transfer rate curves stabilizing around 900,000 KB/s with minor fluctuations. We simulate the disconnection of NIC 2 during the 1st–2nd minute and the 4th–5th minute. The average transition interval from fault detection to task migration across surviving networks remains below 200 milliseconds. This is achieved through Nezha's packet-level task allocation strategy, which ensures the continuity of atomic operations. During these periods, the unaffected TCP network (red curve) takes over all allreduce tasks and switches to a single-rail network state. The results demonstrate that Nezha effectively enables other member networks in the multi-rail setup to take over tasks when a member network fails, ensuring continuous operation.  Moreover, while noticeable performance fluctuations occasionally occur during the allreduce process, they are not caused by NIC failures. These fluctuations appear synchronously on both NICs, indicating that they do not disrupt load balancing.


\section{Experimental Evaluation}
In this section, we investigate the performance characteristics of Nezha. Firstly, we introduce the software and hardware configurations of the experiment as well as the experimental setup. Subsequently, we provide a comprehensive overview using benchmark-level evaluation and application-level evaluation. Through these experiments, we aim to gain a deeper understanding of the impact of different configurations on the performance of the Nezha system and provide guidance for using allreduce on multi-rail networks.

\subsection{Experimental Setup}
{\bfseries Testbeds. }
We evaluated Nezha's performance gains in multi-rail networks using various network combinations on an 8-node local test platform. Additionally, we rented 16 GPU nodes on cloud server clusters and 128 CPU nodes on supercomputing systems to assess its scalability in homogeneous multi-rail networks. Each local node features an Intel Xeon Gold 6230R processor, two V100 GPUs, three Ethernet NICs, one IB NIC, and one TH NIC, all connected via PCIe 3.0 x16. Each cloud node includes an Intel Xeon Gold 5318Y processor, one V100 GPU, one Ethernet NIC, and one IB NIC. Each supercomputer node is equipped with an AMD EPYC 7452 processor, one Ethernet NIC, and one IB NIC. Details of the NIC configurations for the three clusters are shown in Table \ref{tab2}.

\begin{table}[ht]
\caption{Configurations of NICs.}
\centering
\begin{tabular}{cccccc}
\hline
 Cluster&NIC & Net & RDMA & Bandwidth \\
\hline
\multirow{3}{*}{{Local}}  &
MCX623106AN & Eth & No & 100Gbps\\
&ConnectX-5 & IB & Yes& 100Gbps \\
&TH NIC & TH & Yes& 128Gbps \\
\hline
\multirow{2}{*}{{Cloud}}  &
MCX623106AN & Eth & No & 100Gbps\\
&ConnectX-5& IB & Yes& 100Gbps \\
\hline
\multirow{2}{*}{{Supercomputer}}  &
BCM5720 & Eth & No & 1Gbps\\
&ConnectX-3& IB & Yes& 56Gbps \\
\hline
\end{tabular}
\label{tab2}
\end{table}

{\bfseries Implementation. }Nezha is built on Gloo, with an additional 7,000 lines of code. Gloo and NCCL are the two major communication libraries supported by default in current mainstream training \cite{deepspeed,shoeybi2019megatron} and inference \cite{kwon2023efficient,sglang} frameworks. Nezha is fully compatible with Gloo's application ecosystem, inheriting its support for numerous frameworks and further enhancing its capabilities.

{\bfseries Workloads. }We use the Gloo's allreduce benchmark, which performs 10000 consecutive allreduce operations for a specified data volume in a distributed environment and reports the average latency and throughput, with all computational tasks executed by the CPU. Then, Nezha is integrated into the Horovod \cite{sergeev2018horovod} framework to facilitate data-parallel training of DNN models such as VGG-11 \cite{VGG} and AlexNet \cite{AlexNet} on ImageNet ILSVRC2012 \cite{deng2009imagenet}. Finally, Nezha is integrated into the vTrain \cite{vtrain} framework to facilitate hybrid parallel training of the large language model GPT-3 \cite{gpt} on BookCorpus \cite{BookCorpus}.

{\bfseries Metrics. }We employ the average latency of allreduce operations per node and the throughput, defined as the amount of data processed per second, as performance metrics in the benchmark tests. Additionally, for model training, we measure the number of samples processed per second per node and the time taken to complete one iteration.

{\bfseries Baselines. }In the benchmark-level evaluation, we compare Nezha with existing multi-rail network data distribution strategies, MPTCP \cite{ecf} and MRIB \cite{2004Building}. In the application-level evaluation, we compare Nezha with mainstream communication libraries, MPI \cite{gabriel2004open}, NCCL \cite{nvidia2017nccl} and Gloo \cite{meta2017gloo}. MPTCP and MRIB are not integrated into the communication library, hence their exclusion from the application-level evaluation. Nezha incorporates three baseline networks: GLEX, SHARP, and TCP. It facilitates the combination of these baseline networks but does not support the homogeneous combination of GLEX and SHARP networks due to hardware configuration constraints. Each node is equipped with a unique set of devices for SHARP and GLEX networks, and using additional sets of these devices on a single node would lead to hardware conflicts. When evaluating the performance improvement ratio of multi-rail networks over single-rail networks, we consider the performance of the most efficient network within the multi-rail setup in a single-rail context as the baseline.

\subsection{Benchmark Level Evaluation}
\label{benche}

In this subsection, We employ Gloo's benchmark to evaluate the performance of Nezha under homogeneous and heterogeneous network configurations across 4 and 8 nodes.


\subsubsection{Homogeneous multi-rail networks}

The latency and throughput of homogeneous multi-rail networks are shown in Fig.\ref{hetnet}. For small-scale data (2KB to 128KB), MRIB and MPTCP strategies result in higher latency than single-rail networks. In this case,  the link bandwidth is not saturated, and multi-rail networks do not improve data transfer efficiency and incur additional overhead from data synchronization and resource contention. Nezha addresses this by using a single-rail network in the \textbf{cold start} state, reducing startup overhead by at least 15\% compared to MRIB and MPTCP. Nezha dynamically adjusts the threshold for transitioning from the \textbf{cold start} to the \textbf{hot start} state in response to changes in node scale. For TCP-TCP networks, the threshold is 256KB for 4 nodes and decreases to 128KB for 8 nodes. This adjustment is due to the increased likelihood of link bandwidth saturation with a higher number of nodes.


\begin{figure*}[t!]
    \setlength{\abovecaptionskip}{1pt}
 \centering
  \begin{subfigure}[h]{\textwidth}
    \centering
    \begin{subfigure}[h]{0.48\textwidth}
      \includegraphics[width=\textwidth]{figure/aslatency.png}
      \label{aslatency}
    \end{subfigure}  
    \begin{subfigure}[h]{0.48\textwidth}
      \includegraphics[width=\textwidth]{figure/asspeed.png}
      \label{asspeed}
    \end{subfigure}
        \setlength{\abovecaptionskip}{-1pt}
    \caption{TCP-SHARP multi-rail networks}
  \end{subfigure}
  \begin{subfigure}[h]{\textwidth}
    \centering
    \begin{subfigure}[h]{0.48\textwidth}
      \includegraphics[width=\textwidth]{figure/aglatency.png}
      \label{GTlatency}
    \end{subfigure}  
    \begin{subfigure}[h]{0.48\textwidth}
      \includegraphics[width=\textwidth]{figure/agspeed.png}
      \label{agspeed}
    \end{subfigure}
        \setlength{\abovecaptionskip}{-1pt}
    \caption{TCP-GLEX multi-rail networks}
  \end{subfigure}    
  \caption{Latency and throughput of performing allreduce on heterogeneous dual-rail networks.}
  \label{hetnet3}
\end{figure*}

\begin{figure}[!t]
  \centering
  \resizebox{\linewidth}{!}{\includegraphics{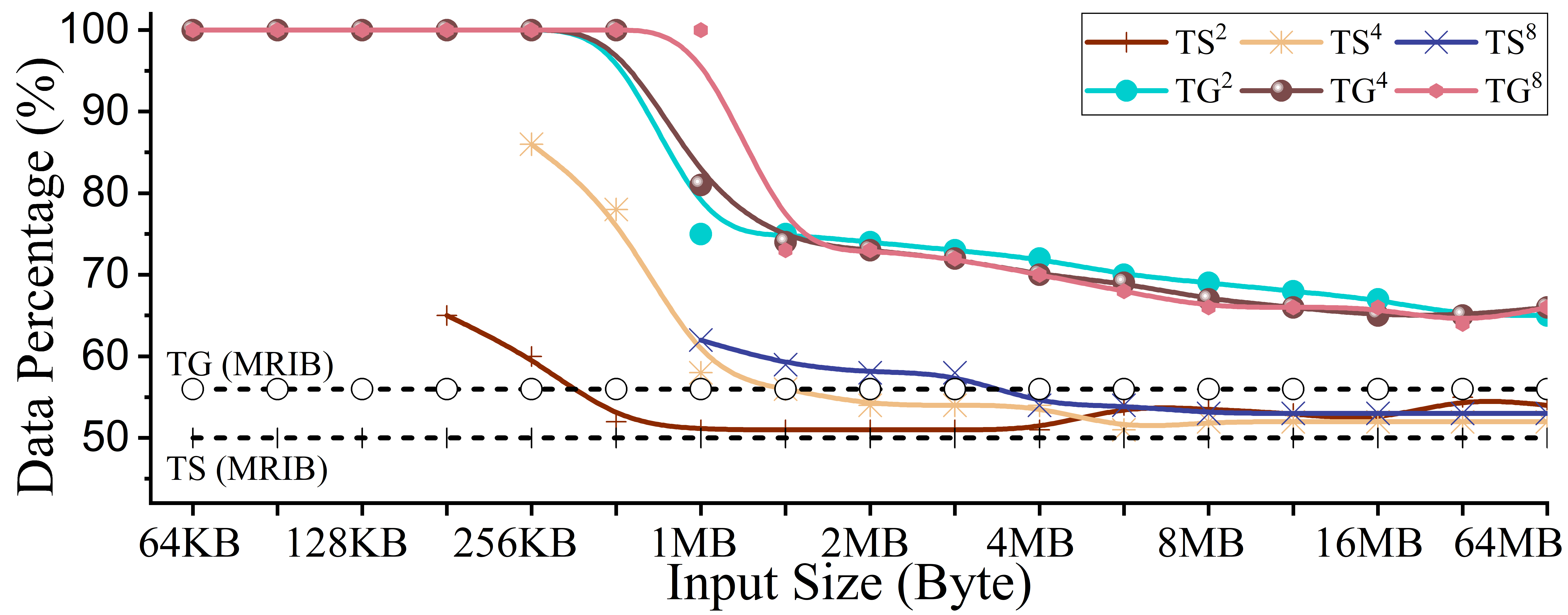}}
  \caption{Data allocation ratio in dual-rail networks. TS and TG respectively represent the data proportion allocated to non TCP networks in TCP-SHARP and TCP-GLEX. The superscript on data indicates the number of nodes.}
  \label{ratio}
\end{figure}

When processing large-scale data (512KB to 64MB) across 4 nodes, MRIB, MPTCP, and Nezha achieve maximum throughput improvements of 84\%, 58\%, and 84\%, respectively, compared to the baseline single-rail network. At this scale, MRIB's performance closely matches Nezha's due to its proportional data allocation based on NIC bandwidth. In homogeneous multi-rail networks, data is evenly distributed across channels, and Nezha follows a similar strategy after Load Balancer assessment. In contrast, MPTCP's performance is somewhat lower because it fragments data into multiple segments, assigning them to idle channels, which introduces additional communication and processing overheads.

When the number of nodes is scaled up to 8, MRIB, MPTCP, and Nezha achieve maximum throughput improvements of 87\%, 50\%, and 87\%, respectively, compared to the baseline single-rail network. The performance gains of MRIB and Nezha increase with more nodes, highlighting the effectiveness of multi-rail networks in managing increased communication bandwidth demands. However, MPTCP's performance improvement ratio declines due to higher communication startup overhead and increased performance loss from frequent data segmentation.

\subsubsection{Heterogeneous multi-rail networks}


The latency and throughput data of heterogeneous multi-rail networks are depicted in Fig.\ref{hetnet3}. In experimental data from TCP-SHARP and TCP-GLEX, common trends are observed. For small-scale data transfers from 2KB to 128KB, MRIB and MPTCP increase the latency of multi-rail networks to near that of single-rail TCP networks, due to synchronization mechanisms that cause low-latency networks to wait for others. Nezha, in handling small-scale data, exploits the low-latency advantages of RDMA by assigning all allreduce tasks to a single high-speed rail.

\begin{figure*}[!t]
  \centering
  \resizebox{\textwidth}{!}{\includegraphics{figure/newsp.png}}
  \caption{Average model training speed, where N represents the number of nodes and bs represents the batch size. Except for TCP (Gloo), TCP (MPI) and TCP (NCCL), the remaining networks are provided by Nezha.}
  \label{horo}
\end{figure*}

\begin{figure}[!t]
  \centering
  \resizebox{\linewidth}{!}{\includegraphics{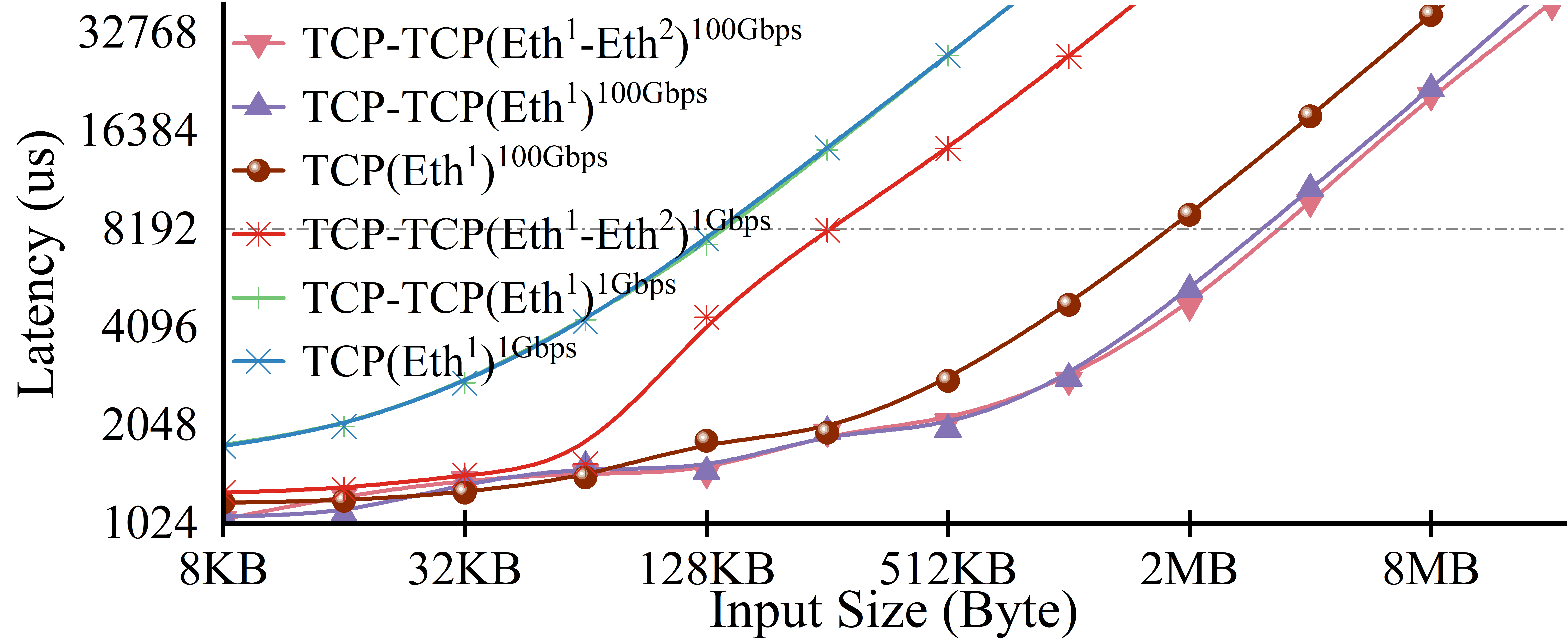}}
  \caption{Latency of allreduce on multi-rail and single-rail networks under different NICs configurations.}
  \label{cpu2}
\end{figure}

In scenarios involving large-scale data (ranging from 512KB to 64MB) across 4 nodes, MRIB and MPTCP's effectiveness in heterogeneous multi-rail networks is diminished. In TCP-GLEX networks, MPTCP achieves only an 8\% throughput increase over the GLEX single-rail network for 2MB data, while MRIB sees an 8\% performance improvement for data over 16MB. MRIB's data allocation, based on NIC bandwidth alone, overlooks the performance gap between RDMA and TCP; MPTCP's strategy fails to account for the allreduce overhead of data blocks assigned to idle networks, resulting in TCP networks' slow processing of trailing data blocks and causing other networks to wait. Nezha excels in enhancing performance in heterogeneous multi-rail networks, boosting throughput by up to 52\% and 46\% over single-rail networks in TCP-SHARP and TCP-GLEX experiments, respectively. Compared to MRIB, Nezha's maximum throughput increase is 41\%, and compared to MPTCP, it is 80\%. Performing the allreduce operation on 8 nodes, Nezha increases the maximum throughput gain for SHARP and GLEX networks to 63\% and 47\%, respectively, demonstrating that higher network stress leads to significant performance gains for Nezha.

\subsubsection{Data allocation ratio between multi-rail networks}
\label{dataall}

Fig.\ref{ratio} illustrates the data allocation ratios assigned by Nezha and MRIB in heterogeneous network combinations. MPTCP, not allocating data based on a ratio for each network, is excluded. Observations include: Firstly, MRIB retrieves bandwidth information of each network during initialization and assigns a fixed data processing ratio to each channel accordingly. However, due to varying network scalability, Nezha dynamically adjusts data distribution to maintain load balance, with ratios varying by node count. Secondly, data distribution in heterogeneous networks should reflect the differential impact of data size across networks and remain dynamic, not static.

\subsubsection{Computation-communication trade-offs in Nezha}
\label{fact}

In subsection \ref{cpure}, we demonstrate that the performance of allreduce is constrained by CPU resource allocation.Indeed, the bottleneck of CPU resources primarily arises from the mismatch between network hardware and compute hardware performance. To substantiate this, we test the latency of the allreduce operation using a TCP-TCP combination with 2 identical NICs per node (denoted as TCP-TCP(Eth$^1$-Eth$^2$)), the latency with the virtual TCP-TCP combination using 1 NIC per node (TCP-TCP(Eth$^1$)), and the latency with a single-rail TCP network using 1 NIC per node (TCP(Eth$^1$)). Fig.\ref{cpu2} illustrates that with 1Gbps NICs, the latency of TCP-TCP(Eth$^1$) does not fall below that of TCP(Eth$^1$), indicating that network bandwidth constitutes the performance bottleneck for allreduce operations under these conditions. In contrast, with 100Gbps NICs, the latency of TCP-TCP(Eth$^1$) is only slightly higher than that of TCP-TCP(Eth$^1$-Eth$^2$) for data exceeding 2MB, yet it remains significantly lower than the latency of TCP(Eth$^1$). This suggests that CPU resources become the primary constraint on performance, thereby highlighting the advantages of TCP-TCP networks in achieving higher CPU utilization for allreduce tasks.

\begin{figure*}[t!]
  \centering
  \resizebox{\textwidth}{!}{\includegraphics{figure/horobilinew2.png}}
  \caption{Latency of allreduce on single-rail and dual-rail networks when training AlexNet on 4 nodes. The proportion of data allocated to each network is recorded in parentheses, and Opt. indicates the optimal data allocation ratio in Fig.\ref{ratio}.}
  \label{horobil}
\end{figure*}

\begin{figure}[t!]
  \centering
  \resizebox{\linewidth}{!}{\includegraphics{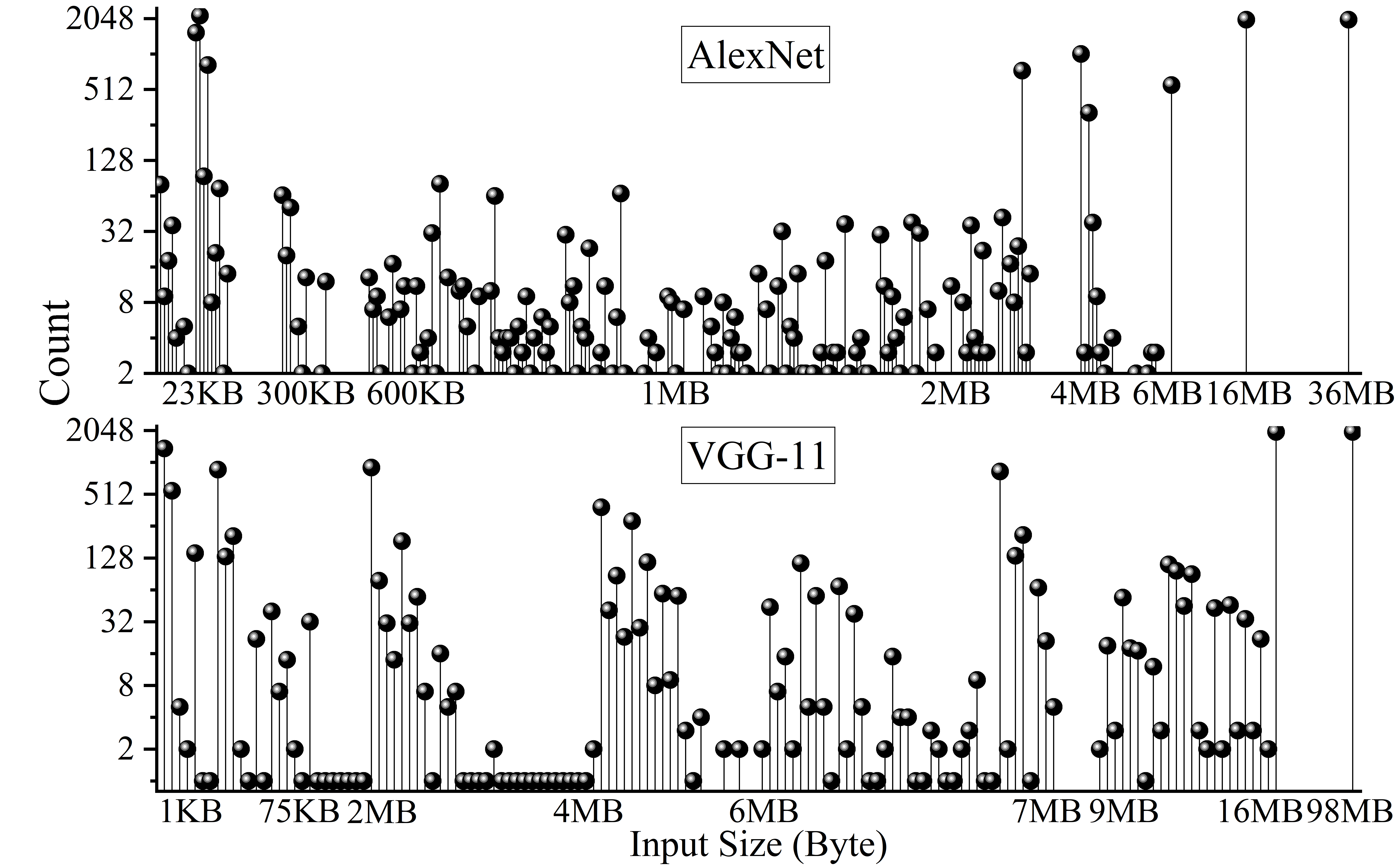}}
  \caption{Allreduce count and data size involved per epoch during models training.}
  \label{para}
\end{figure}

\subsection{Application Level Evaluation}
\label{ap}

We employ Nezha to facilitate communication during the training of AlexNet and VGG-11 models across 4 and 8 nodes. The training speed is documented in Fig.\ref{horo}, revealing the subsequent observations. Firstly, homogeneous dual-rail networks offer superior scalability compared to the single-rail network. When training the VGG-11 model with a batch size of 64 on 4 and 8 nodes, Nezha's TCP-TCP network demonstrates performance improvements of 19.9\% and 50.4\%, respectively, over Gloo's TCP network. The dual-rail networks enhance the parallelism between computation and communication, thereby mitigating transmission pressure that escalates with the node count increment.

Secondly, the performance acceleration observed in multi-rail networks varies across different models. For example, when training AlexNet and VGG-11 models on 8 nodes (with a batch size of 32 and PCIe3.0x16), the TCP-TCP networks yield performance enhancements of 70.1\% and 52.1\%, respectively, compared to NCCL's TCP network. 


Thirdly, the performance gains that multi-rail networks bring to the network with better data transfer capabilities are more modest. When training AlexNet and VGG-11 models on 8 nodes, the TCP-GLEX networks achieve maximum training speed improvements of only 11.6\% and 11.1\%, respectively, compared to the GLEX network, while the TCP-SHARP networks achieve up to a 20.1\% improvement over the SHARP network. This discrepancy is due to the significant difference in data transfer rates between the GLEX network and others. According to Fig. \ref{rho}, the performance gain obtained by the optimal network decreases as the real-time efficiency ratio $\rho(S)$ between networks increases.

Lastly, the PCIe bus bandwidth does not limit the performance of multi-rail networks. We downgrade  the bus protocol of each device from PCIe 3.0x16 to PCIe 2.0x16. As shown in Fig. \ref{horo}, Nezha's performance advantage in multi-rail networks remains unaffected. This indicates that Nezha can be extensively deployed across clusters without requiring stringent hardware specifications.


\subsubsection{Communication characteristics of models}
\label{chrac}


We employ the Control Module to record the frequency of allreduce operations and the volume of data exchanged by each node during model training, as depicted in Fig.\ref{para}. Communication activities in AlexNet primarily involve data sizes below 4MB, while VGG-11 engages in intensive communication across the data size range of 2MB to 16MB. This observation elucidates the varying acceleration effects of multi-rail networks for different models, as distinct networks demonstrate unique overhead characteristics when transmitting diverse data volumes. Furthermore, in data-parallelism training, Nezha can further support the training of larger models without any modifications. From a communication perspective, the differences between models lie solely in the size of the parameters involved in communication and the communication frequency used to transmit various sizes of parameters.

\subsubsection{Multi-NIC vs. Single-NIC}
\label{mulsin}

We conduct further investigation into the latency of allreduce  on 4 nodes during the training of the AlexNet model. To discern disparities between networks in multi-rail and single-rail configurations, we establish the following comparison groups: single-rail network, multi-rail networks with data allocation based on load balancing, and multi-rail networks with data allocation in a 99:1 ratio. The outcomes are delineated in Fig. \ref{horobil}. With computational resources held constant, the transmission latency of member networks within multi-rail networks is not interfered with by other member networks. For instance, when considering TCP networks, whether in heterogeneous TCP-GLEX networks, TCP-SHARP networks, or homogeneous TCP-TCP networks, the latency varies between 1024us and 8192us when 1\% of the data is allocated. Additionally, Nezha's data distribution strategy ensures balanced latency across each member network. Even in the presence of performance fluctuations within the system, the average scheduling error is maintained within 9.3\%.

Besides, due to the constraints of thread synchronization and the limitations of computational resources, member networks in multi-rail networks experience a reduction in transmission rate when performing allreduce compared to single-rail network configurations. The extent of this loss is influenced by the sensitivity of the network to the number of CPU cores. This conclusion is drawn from the latency comparison between GLEX (99\%), SHARP (99\%), TCP (99\%), GLEX, SHARP, and TCP networks, where the average latency increase is 17.5\% for GLEX(99\%), 15.6\% for SHARP(99\%), and 9.7\% for TCP(99\%). However, the proportion of performance degradation in member networks decreases as the scale of nodes increases: when testing the same configuration with 8 nodes, the performance degradation ratios for GLEX (99\%), SHARP (99\%), and TCP (99\%) compared to single-rail networks are 15.7\%, 13.4\%, and 8.3\%, respectively.


\begin{figure*}[t!] 
\centering
\begin{minipage}{0.75\textwidth}
\centering
\resizebox{1\textwidth}{!}{ 
\begin{tabular}{ c|*{5}{c}|*{5}{c} }
\multirow{2}{*}{{Model}}  & \multicolumn{5}{c|}{{4 Node}} & \multicolumn{5}{c}{{6 Node}}   \\
 & {\textit{G$_1$N$_1$}}  & {\textit{G$_1$N$_2$}} & {\textit{G$_1$N$_3$}}& {\textit{G$_2$N$_1$}} & {\textit{G$_2$N$_2$}} & 
{\textit{G$_1$N$_1$}}  & {\textit{G$_1$N$_2$}} & {\textit{G$_1$N$_3$}}&  
{\textit{G$_2$N$_1$}} & {\textit{G$_2$N$_2$}}  \\
\hline
Alex$_{32}$ &215.7 & 319.0 (1.48) & 376.8 (1.74)& 420.3 (1.95) & 516.5 (2.39) &283.9 & 415.4 (1.46) & 462.1 (1.63)&559.7 (1.97)  & 731.7 (2.58) \\
\hline
Alex$_{64}$ &426.6  & 612.5 (1.44) & 753.2 (1.77)&846.3 (1.98) & 1075.5 (2.52) &569.3  & 835.7 (1.47) & 929.7 (1.63)&1132.4 (1.99) & 1471.9 (2.58) \\
\hline
VGG$_{32}$ &114.8  & 154.8 (1.35) & 179.2 (1.56)&207.0 (1.80) & 242.3 (2.11) &147.2  & 215.9 (1.47) & 244.5 (1.66) &286.5 (1.95)& 366.5 (2.49) \\
\hline
VGG$_{64}$ &253.2 & 303.7 (1.20) & 362.5 (1.43) &395.7 (1.56) & 508.5 (2.01) &292.1 & 412.7 (1.41) & 456.7 (1.56)&569.4 (1.95)  & 722.4 (2.47) \\
\end{tabular}
}
\caption{AlexNet and VGG-11 training speeds. Values in parenthesis indicate how many times faster the system is compared to the baseline single GPU and single NIC system (G$_1$N$_1$). Alex$_{32}$ signifies AlexNet model training with a batch size of 32. G$_2$N$_2$ denotes the configuration of 2 GPUs and 2 NICs per node.}
\label{tab3}
\end{minipage}
\hfill 
\begin{minipage}{0.24\textwidth}
\centering
\resizebox{\textwidth}{!}{\includegraphics{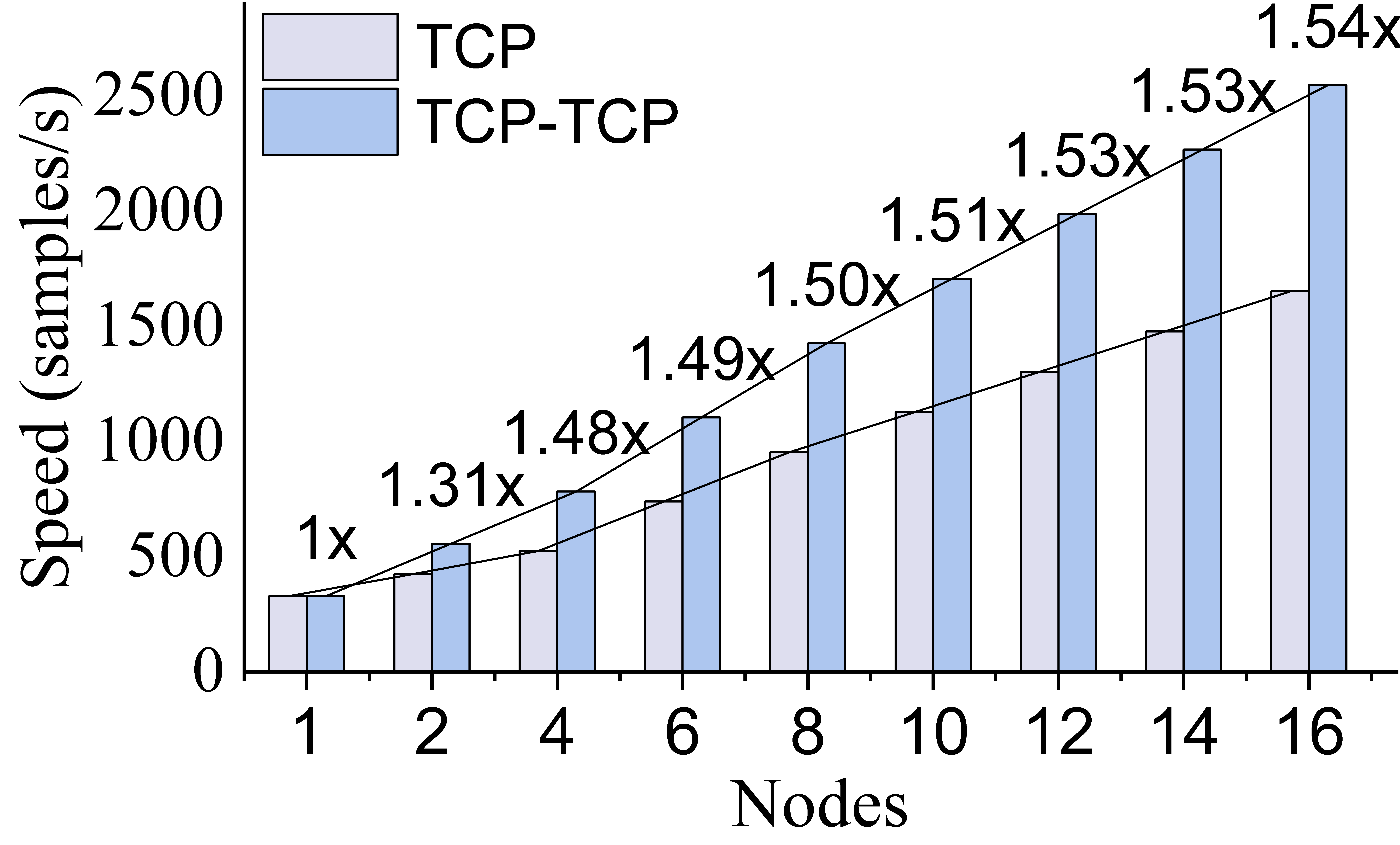}}
\caption{AlexNet training speeds on multiple nodes.}
\label{scal2}
\end{minipage}
\end{figure*}

\subsubsection{Multi-GPU vs. Multi-NIC}

Currently, the majority of cluster configurations incorporate NICs that are compatible with the TCP protocol, enabling the broad deployment of Nezha to provide multi-TCP network services. To assess the performance enhancements achieved by increasing Ethernet NICs and integrating GPUs, we conduct comparative analyses aimed at offering insights into the deployment of multi-TCP networks on cloud servers. Our experiments structure the following comparison groups across 4 and 6 nodes: single GPU with single NIC (serving as the baseline), single GPU with multiple NICs, dual GPUs with single NIC, and dual GPUs with multiple NICs. The corresponding experimental data is presented in Fig. \ref{tab3}, where model training speed is the primary metric for performance, with the values in parentheses representing the performance ratio relative to the baseline.

Our observations yield several insights: Firstly, augmenting Ethernet NICs improves system performance and holds significant practical value. Notably, on 6 nodes, the performance of {\textit{G$_1$N$_2$}} (1 GPUs and 2 NICs per node) is comparable to that of {\textit{G$_2$N$_1$}} on 4 nodes, suggesting that incorporating 4 Mellanox 100Gbps NICs in our 6-node cluster brings a performance enhancement equivalent to adding 1 V100 GPU. Secondly, multi-rail networks complement multi-GPU configurations, as {\textit{G$_2$N$_2$}} brings over 30\% performance improvement compared to {\textit{G$_2$N$_1$}}. Thirdly, multi-rail networks can gain a greater advantage in environments where communication is the bottleneck. On 6 nodes, {\textit{G$_1$N$_3$}} achieves only around a 10\% increase compared to {\textit{G$_1$N$_2$}}, while the performance improvement of {\textit{G$_2$N$_2$}} compared to {\textit{G$_2$N$_1$}} exceeds that of {\textit{G$_1$N$_2$}} compared to {\textit{G$_1$N$_1$}}. This implies that on clusters equipped with more powerful GPUs, such as H100 \cite{hgpu} or B100 \cite{bgpu}, Nezha can substantially enhance training efficiency. Fourthly, dual-rail networks demonstrate superior scalability compared to the single-rail network, as the performance improvement of {\textit{G$_x$N$_2$}} over {\textit{G$_x$N$_1$}} on 6 nodes exceeds that on 4 nodes. Lastly, disregarding limitations of other system resource, the configuration of {\textit{G$_1$N$_3$}} can be interpreted as one GPU paired with a combination of a 200Gbps and a 100Gbps Ethernet NICs, highlighting the potential of heterogeneous NICs to enhance multi-TCP networks.


\subsubsection{Scalability analysis}


\begin{table}[t!]
\centering
\caption{Training Setup for GPT-3 Model Training}
\label{tabfinal}
\begin{tabular}{lcccc}
\toprule
Training Configuration & N=16 & N=32 & N=64 & N=128 \\
\midrule
TP/DP/PP  & 2/2/8 & 2/4/8 & 2/8/8 & 2/16/8 \\
Batch Size & 128 & 512 & 512 & 512 \\
\bottomrule
\end{tabular}

\vspace{0.5em}
\begin{minipage}[t]{0.9\linewidth}
\footnotesize
TP (Tensor Parallelism), and DP (Data Parallelism), PP (Pipeline Parallelism) specify the degree of each parallelism type in the 3D configuration. The calculation graph is generated based on 2 V100 GPUs per node.
\end{minipage}
\end{table}

We conduct tests on the cloud server cluster with varying numbers of nodes to evaluate the average training speed of the AlexNet model using Nezha's TCP-TCP multi-rail networks and Gloo's TCP single-rail network as the Horovod communication backend. Fig. \ref{scal2} presents the relevant data, comparing the performance improvement ratio of the TCP-TCP network to the TCP network. The results show that as the number of nodes increases, the performance improvement ratio of the TCP-TCP network over the single-rail network also increases, further confirming the superior scalability of Nezha's TCP-TCP network.

\begin{figure}[t!]
 \centering
  \begin{subfigure}[h]{0.24\textwidth}
    \includegraphics[width=\textwidth]{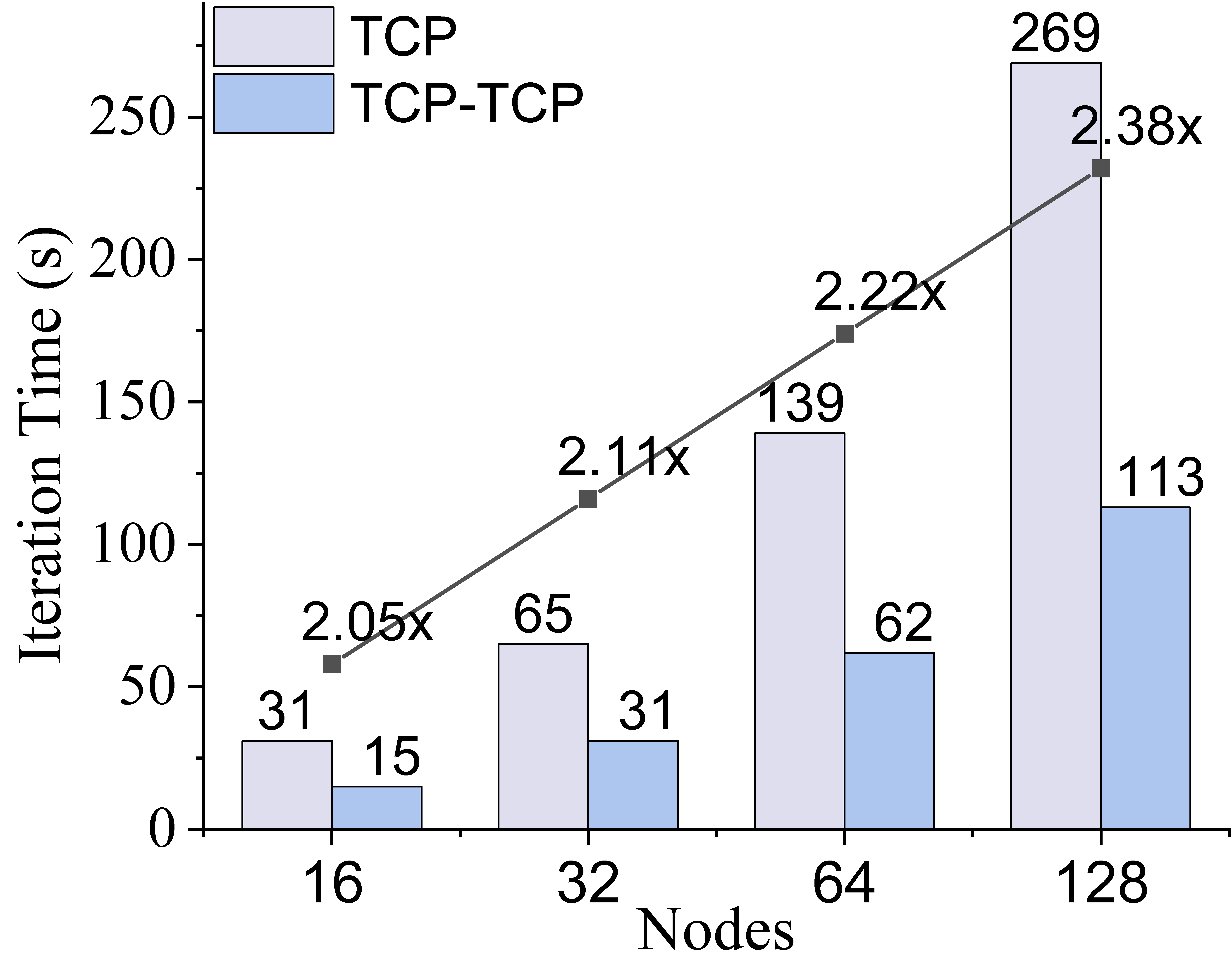}
      \caption{GPT-2.7B}
    \label{gpt1}
  \end{subfigure}  
  \begin{subfigure}[h]{0.24\textwidth}
    \includegraphics[width=\textwidth]{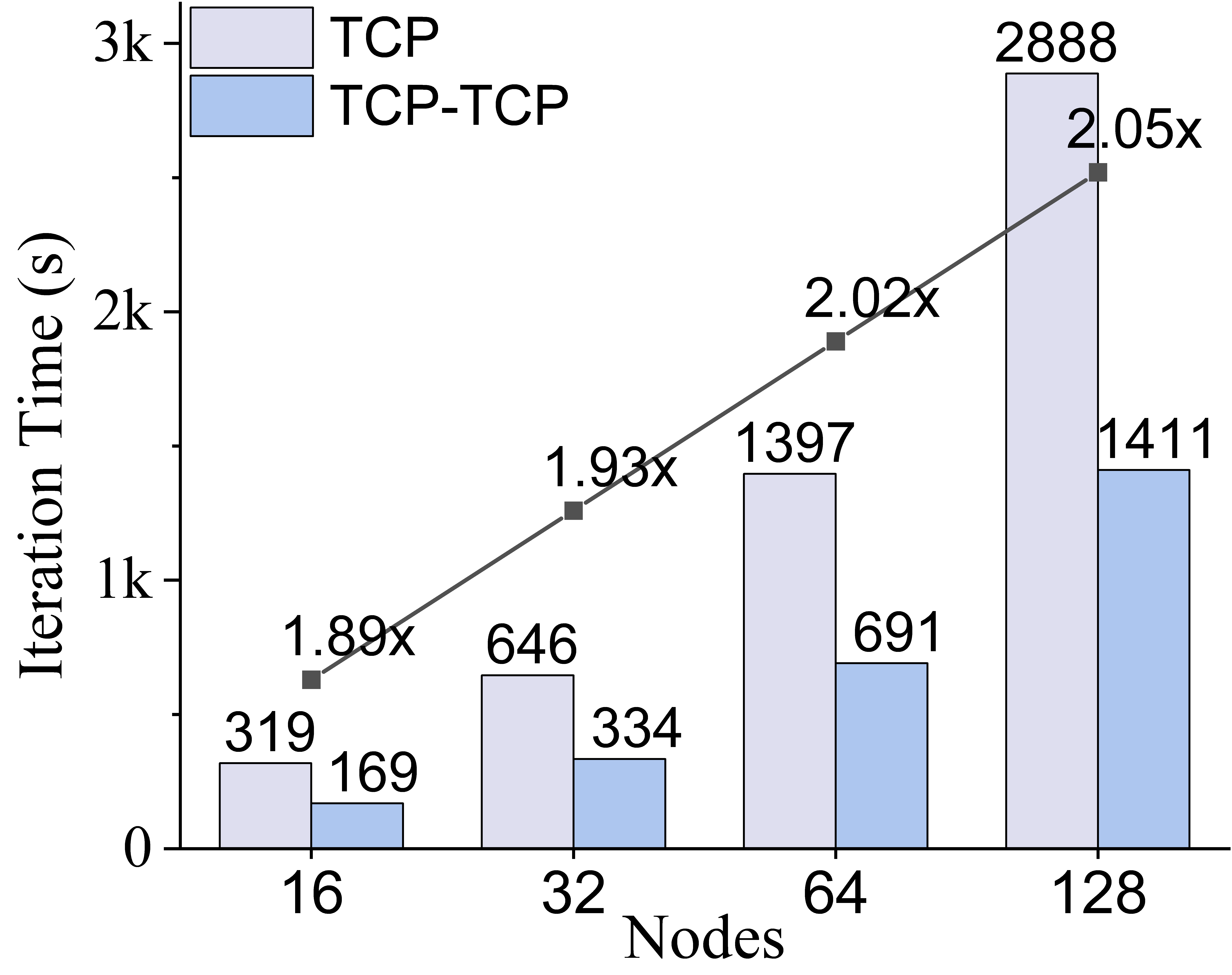}
          \caption{GPT-30B}
    \label{gp2}
  \end{subfigure}  
  \caption{The training iteration time of GPT on multiple nodes using Ring allreduce algorithm.}
  \label{gpt}
\end{figure}
\begin{figure}[t!]
 \centering
  \begin{subfigure}[h]{0.24\textwidth}
    \includegraphics[width=\textwidth]{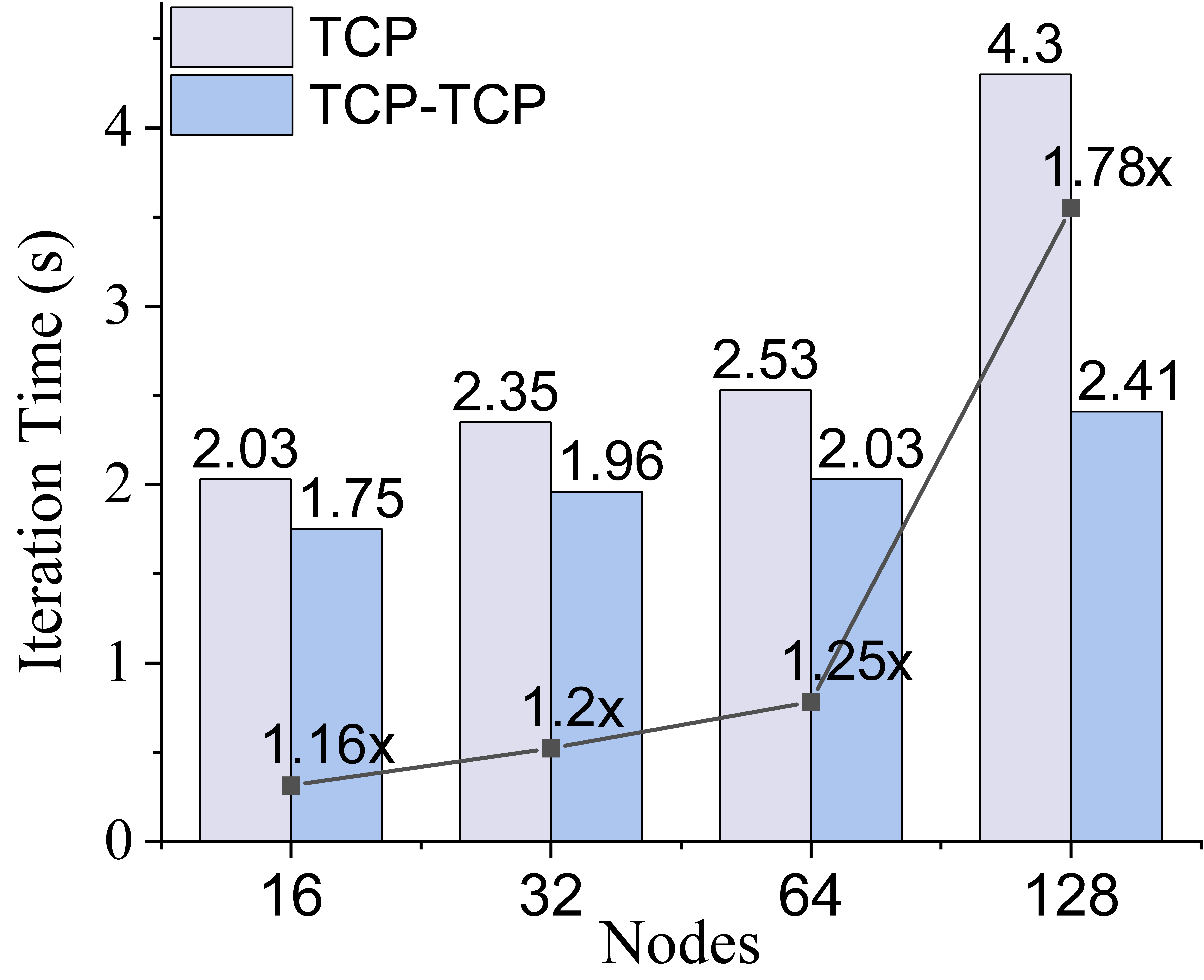}
      \caption{GPT-2.7B}
    \label{gpt3}
  \end{subfigure}  
  \begin{subfigure}[h]{0.24\textwidth}
    \includegraphics[width=\textwidth]{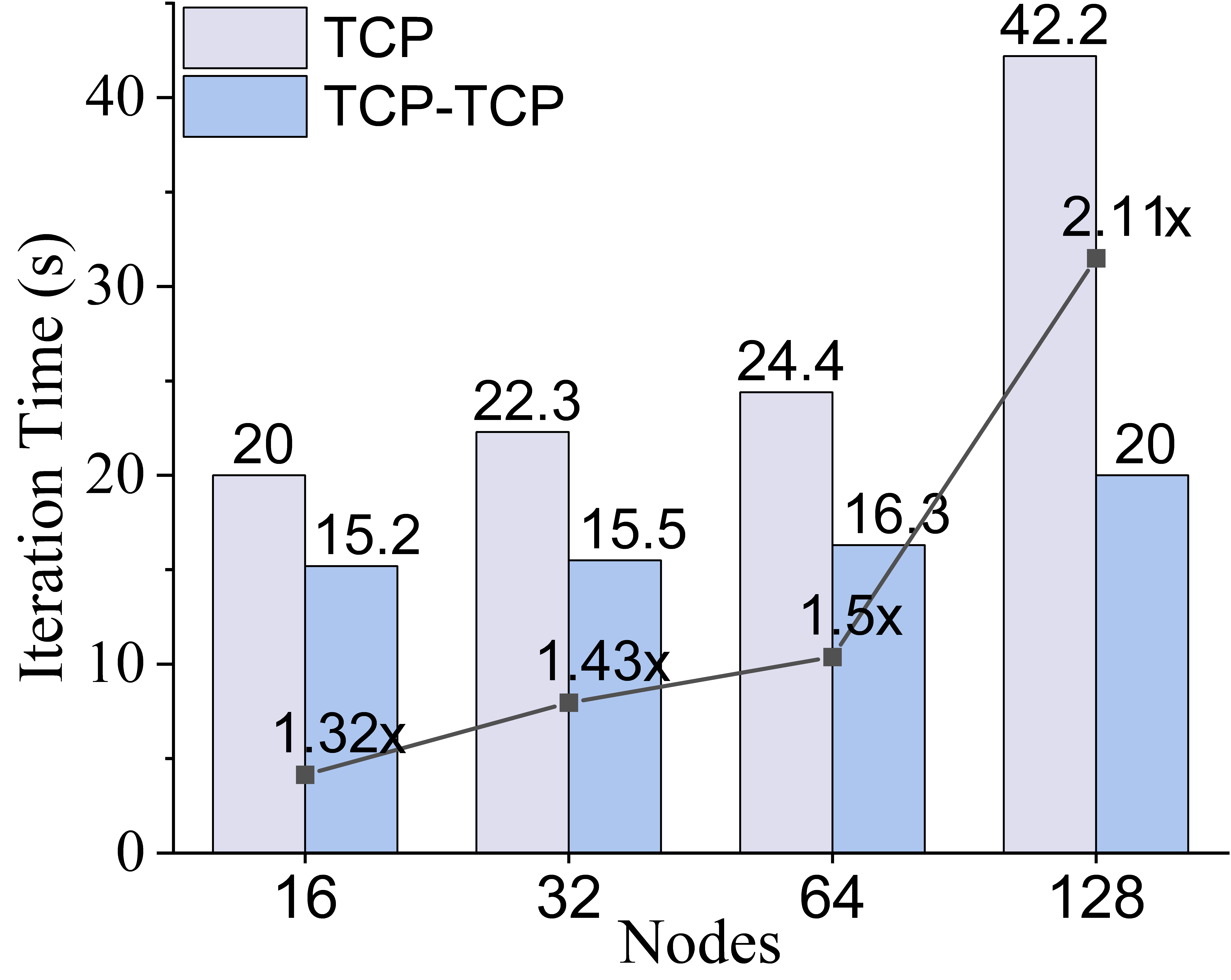}
          \caption{GPT-30B}
    \label{gp4}
  \end{subfigure}  
  \caption{The training iteration time of GPT on multiple nodes using Ring\_Chunked allreduce algorithm.}
  \label{gptt}
\end{figure}


To evaluate the scalability of Nezha for larger language models, we integrate Nezha into the vTrain \cite{vtrain} framework to test the training overhead of GPT-3 on large-scale supercomputing nodes. vTrain utilizes CPU to virtually execute CUDA computation graphs. It mirrors the overhead of each computation operation by reading a pre-measured overhead table from a GPU. The communication initiation timing and data packet sizes during training are consistent with real training scenarios. Due to the significant performance difference between the cluster's 56Gbps IB NICs and 1Gbps Ethernet NICs, we throttle the IB NICs to 1Gbps for these tests. We measure the average training overhead per node for GPT-3 (2.7B and 30B) across various node scales, with parallel configurations and global batch sizes as shown in Table \ref{tabfinal}. At 16 nodes, setting the batch size to 512 causes out-of-memory errors, so we reduce it to 128. 
When training GPT-3 (30B) using Gloo's single-rail TCP network as the backend, we encounter Segmentation Fault errors. This is due to gradients data packets exceeding 1GB, which causes the NICs to crash. Nezha does not experience this issue because it distributes the data evenly across two network links. To resolve this and ensure a fair comparison, we split data packets larger than 1GB into multiple 256MB packets.
Using the Ring allreduce algorithm, the training overhead across different node scales is shown in Fig. \ref{gpt}. As the number of nodes increases, communication overhead rises significantly. NaZha's dual-rail TCP allreduce outperforms Gloo's allreduce. As the number of nodes increases, the efficiency gap widens. At 128 nodes, the training efficiency improves by 2.38x, exceeding the theoretical value of 2x. This is mainly because the dual-rail network can effectively reduce packet collisions, lower transmission delays, and decrease retransmission rates in bandwidth-limited scenarios, thereby significantly enhancing communication efficiency.

Additionally, following Gloo's official recommendations, we test the Ring\_Chunked allreduce algorithm (which splits large data packets and pipelines their transmission). The results, shown in Fig. \ref{gptt}, indicate that Ring\_Chunked allreduce effectively reduces training iteration time. At node scales of 64 and below, iteration overhead does not spike with increasing nodes, and Nezha's efficiency improvement over Gloo grows modestly. However, at 128 nodes, communication overhead surges, leading to a significant increase in iteration overhead and a marked rise in Nezha's efficiency improvement over Gloo.

\section{Related Work and Discussion}
The related work encompasses three aspects: optimizing software-hardware collaborative communication, enhancing allreduce acceleration, and exploring multi-rail networks.

\textbf{Optimizing software-hardware collaborative communication:} 
Early studies, such as QsNet \cite{946704} and Portals \cite{6687397}, focus on network devices that offload basic data operations from the CPU. SwitchML \cite{sapio2021scaling} later introduces a system that offloads data aggregation to the switch, addresses floating-point data storage and computation, as well as packet loss. ATP \cite{265053} and A2TP \cite{10.1145/3552326.3587436} further optimize in-network computing by resolving data overflow and congestion. Apart from Nezha, there have been endeavors \cite{9308656, 8924164} to furnish software stack support for SHARP to expedite collective communication operations. However, Nezha does not focus on integrating SHARP and GLEX. They are just mediums for Nezha to investigate the performance of allreduce across various configurations of multi-rail networks.

\textbf{Enhancing allreduce acceleration:} The acceleration of allreduce can be broadly classified into two categories: application framework-based approaches \cite{paszke2019pytorch,chen2015mxnet,abadi2016tensorflow} and communication library-based approaches \cite{nvidia2017nccl,meta2017gloo}. The former encompasses strategies such as scheduling communication operations to enhance the parallelism between computation and communication \cite{wang2022overlap,li2021chimera,DBLP:journals/tsc/WuXLX22}, reducing the number of parameters involved in communication \cite{m2021efficient,renggli2019sparcml}, transmitting data in lower bit formats \cite{alistarh2017qsgd,wen2017terngrad,10.1145/3552326.3567505}, and decreasing communication frequency \cite{pmlr-v202-xu23v,NEURIPS2023_04bd683d,pmlr-v202-guo23g}. The latter concentrates on designing and implementing hardware-compatible communication mechanisms \cite{graham2016scalable,sapio2021scaling}, optimizing allreduce communication algorithms \cite{li2014scaling, gibiansky2017bringing, sanders2009two}, and aligning logical topologies with physical topologies \cite{sanghoon2023logical,wang2021impact}. Nezha belongs to the communication library-based approach, making multi-rail network architectures visible and schedulable for the field of distributed deep learning. This provides the possibility for further enhancements of the aforementioned related work within multi-rail networks.

\textbf{Exploring multi-rail networks:} MPTCP \cite{ecf} enables devices to use multiple NICs for data streaming, while MRIB \cite{2004Building} employs fixed data allocation ratios across multiple paths for packet transmission and reception. However, neither approach addresses heterogeneous network protocols or supports allreduce operations, making them impractical for distributed deep learning.

Nezha is orthogonal to existing optimization techniques as it can provide communication support for application frameworks such as Horovod \cite{sergeev2018horovod} and PyTorch \cite{paszke2019pytorch}, and integrate topology optimization and communication algorithms into system modules. Therefore, the emergence of this system will bring new opportunities for existing related work.




\section{Conclusion}

Nezha is the inaugural communication library system capable of flexibly scheduling multi-rail networks to support allreduce operations. It supports TCP as well as integrates TH Express-2's GLEX and Mellanox's SHARP. Nezha offers the optimal data allocation scheme for multi-rail networks, optimizing network performance by selecting the network with the lowest latency to handle small data volumes and enabling coordinated operations across multiple networks to transmit large data volumes under load-balancing conditions. It supports both heterogeneous and homogeneous network parallelism for deep learning applications, fully leveraging the system's network configuration to address scalability challenges of allreduce in distributed systems. However, Nezha's primary focus is on enhancing network capabilities, and its benefits may be limited in systems where computational power is the bottleneck. In the future, we plan to evaluate Nezha's performance on more advanced large-scale clusters and assess its capabilities for other collective communication operations.


\ifCLASSOPTIONcompsoc


\ifCLASSOPTIONcaptionsoff
  \newpage
\fi

\bibliographystyle{IEEEtran}
\bibliography{bib3}

\end{document}